\documentclass[aps,prd,twocolumn,showpacs,showkeys,amsmath,amssymb]{revtex4}
\usepackage{graphicx}
\usepackage{dcolumn}
\usepackage{bm}
\usepackage{mathtext}
\usepackage[dvips]{epsfig}
\usepackage{epsfig}
\usepackage{graphics}
\usepackage{subfigure}
\usepackage{mathrsfs}
\usepackage{amssymb,latexsym}

\newcommand{\seq}{\begin{subequations}}
\newcommand{\sen}{\end{subequations}}
\newcommand{\eq}{\begin{eqnarray}}
\newcommand{\en}{\end{eqnarray}}
\newcommand{\ra}{\rangle} 
\newcommand{\la}{\langle} 
\def\shiftleft#1{#1\llap{#1\hskip 0.04em}}
\def\shiftdown#1{#1\llap{\lower.04ex\hbox{#1}}}
\def\thick#1{\shiftdown{\shiftleft{#1}}}
\def\b#1{\thick{\hbox{$#1$}}}
\begin{document}
\title{Electroproduction of the Roper resonance on the proton:\\ 
the role of the three-quark core and the molecular $N\sigma$ component} 

\author{Igor T. Obukhovsky$^1$,
Amand Faessler$^2$,
Dimitry K. Fedorov$^1$,
Thomas Gutsche$^2$, 
Valery E. Lyubovitskij$^2$
\footnote{On leave of absence
from Department of Physics, Tomsk State University,
634050 Tomsk, Russia}
\vspace*{1.2\baselineskip}}

\affiliation{$^1$Institute of Nuclear Physics, Moscow
State University,119991 Moscow, Russia
\vspace*{1.2\baselineskip} \\
\hspace*{-1cm} 
$^2$Institut f\"ur Theoretische Physik,
Universit\"at T\"ubingen,\\
Kepler Center for Astro and Particle Physics, \\
Auf der Morgenstelle 14, D--72076 T\"ubingen, Germany
\vspace*{0.3\baselineskip}\\}

\date{\today}

\begin{abstract}

The Roper resonance is considered as a mixed state of a three-quark core 
configuration and a hadron molecular component $N+\sigma$. Based on 
this ansatz we study electroproduction of the Roper resonance. 
The strong and electromagnetic couplings induced by the quark core are
calculated in the $^3P_0$ model. The contribution of the vector meson cloud 
to the electromagnetic transition is given in the framework of the VMD model. 
Results are compared with the recent JLab electroproduction data.

\end{abstract}



\maketitle

\section{Introduction.}\label{s1}

The structure issue of the lowest lying nucleon re\-so\-nan\-ce 
$N(1440)$ with $J^P = \frac{1}{2}^+$  (the Roper resonance $P_{11}$ or simply 
$R$) has been a long standing problem of hadron physics. One of the indication 
that the inner structure of the Roper is possibly more complicated than the 
structure of the other lightest baryons was obtained some time ago in the 
framework of the constituent quark model (CQM). It was found (see, 
e.g.~\cite{isgur80}) that the observed mass of the Roper resonance is much too 
low and the decay width is too large when compared to the predicted values of 
the CQM.  

The simplest description of the Roper consists of the three-quark $(3q)$ 
configuration $sp^2[3]_X$, i.e. the first ($2S$) radial excitation of the 
nucleon ground state $s^3[3]_X$, but it fails to explain either the large decay 
width $\Gamma_R\simeq$ 300 MeV or the branching ratios for the $\pi N$ 
(55-75\%) and $\sigma N$ (5-20\%) decay channels~\cite{pdg10,sarantsev08}. 
Evaluation of these values in the framework of the CQM is often based on 
the elementary emission model (EEM) with single-particle quark-meson (or 
quark-gamma) couplings $qq\pi$, $qq\sigma$, $qq\gamma$, etc.. The calculation 
of decay widths (or of the  electroproduction cross section at
small virtuality of the photon with $Q^2\simeq\,$0) results in anomalous 
small values. These underestimates can especially be traced to the strict
requirement of orthogonality for the ground ($0S$) and 
excited state ($2S$) radial wave functions of the $N$- and $R$ states belonging 
to the quark configurations with the same spin-isospin ($S=1/2$, $T=1/2$) 
and symmetry ($[3]_{ST}[3]_X$) quantum numbers. To overcome this discrepancy
it is suggested that either the Roper is not an ordinary $3q$ state or the 
"true" transition operators have a more complicated form than the 
single-particle operators used in calculations. 

Quark models with Goldstone boson interactions~\cite{glozman} can explain why 
the mass of the Roper resonance is shifted to the observed value including the
correct level ordering. But these models still fail to get the strong decay
widths and electromagnetic couplings under control.
 
On the experimental side there has been noticeable progress in the experimental
study of the Roper resonance in the last decade. The Roper resonance has been 
studied in $\pi$~\cite{aznauryan09} and $\pi\pi$~\cite{mokeev09} 
electroproduction processes on the proton with the polarized electron beam at 
the JLab (CLAS Collaboration) followed by a combined analysis of pion- and 
photo-induced reactions made by CB-ELSA and the A2-TAPS 
Collaborations~\cite{sarantsev08}. These recent data present new possibilities 
for the study of the lightest baryon resonances. 

Several models for the description of the Roper resonance electroexcitation 
were proposed during the last three decades~\cite{gavella80,li90,close90,%
capstick95,cardarelli97,aznauryan07,buchmann98,dong99,cano98,riska06} 
(see the review~\cite{burkert04} for a detailed discussion). 
Now model predictions can be compared with the new high-quality photo- 
and electroproduction data~\cite{sarantsev08,aznauryan09,mokeev09}, 
and updated versions~\cite{capstick07,romalho10,golli09} of the most
realistic models give a good description of the data at intermediate values of 
1.5$\,\lesssim Q^2\lesssim\,$4 GeV$^2$. 
However, in the "soft" region, i.e. at low values of $Q^2$ 
(0$\,\leq Q^2\lesssim\,$1 - 1.5 GeV$^2$), the data differ qualitatively from 
the theoretical predictions: the experimental helicity amplitude 
$A_{1/2}$ changes sign at $Q^2\approx\,$0.5 GeV$^2$ and it is large 
and negative at the photon point $Q^2=\,$0. Theoretical predictions for
$A_{1/2}$ are large and positive at $Q^2\approx\,$0.5 GeV$^2$ 
and quickly go to a small negative (or zero) value at the photon point.

For pion electroproduction in the resonance region $W\simeq m_R$ the behavior 
of the transverse helicity amplitude $A_{1/2}$ near the photon point 
$Q^2\gtrsim\,$0 is most sensitive to the "soft" component of the resonance 
state, i.e. to the possible contribution of the meson cloud. Electroproduction 
amplitudes in this kinematical region are successfully analyzed in terms of 
the dynamical coupled channel model~\cite{matsuyama07,suzuki09},
which is used at the Excited Baryon Analysis Center (EBAC) at JLab (see, 
e.g.~\cite{kamano10}). The detailed description of the low-$Q^2$ CLAS 
$p\pi^+\pi^-$ data~\cite{fedotov09} was obtained in Ref.~\cite{mokeev09} on 
the basis of JLab-Moscow (JM) model~\cite{aznauryan05,mokeev07} with taking
into account the $\pi\Delta$ channel along with additional contact terms and 
the direct $2\pi$ production.
The contribution of the meson (pion) cloud to the Roper resonance mass was 
recently calculated in Refs.~\cite{Borasoy:2006fk,Djukanovic:2010zz}. 

As a result, there are essentially two comprehensive theoretical approaches to 
the Roper electroproduction on the market. One of them (the coupled channel 
model of the meson cloud~\cite{mokeev09,matsuyama07,suzuki09,kamano10}) is only 
successful in the soft region 0$\,\leq Q^2\lesssim\,$1 GeV$^2$ 
and the other one (the light front (LF) three-quark 
model~\cite{capstick07,aznauryan07} or the covariant quark spectator 
model~\cite{romalho10}) is compatible with data in the hard region 
1.5$\,\lesssim Q^2\lesssim\,$4 GeV$^2$. 

Universal, but more phenomenological approaches whi\-ch pretend to cover both 
regions of $Q^2$ were also suggested (see, e.g. Refs.~\cite{cano98} 
and~\cite{golli09}). In Ref.~\cite{cano98} a $3q+\bar qq$ approach was 
suggested using the $^3P_0$ model~\cite{gavella80} and vector meson 
dominance (VMD) in combination with the EEM. In Ref.~~\cite{golli09}
a generalization of the Cloudy Bag Model (CBM)~\cite{thomas84} was used for
the case of the open inelastic channels $\pi\Delta$ and $\sigma N$
in combination with a phenomenological strong background interaction.

\begin{figure}[hp]
\begin{center}
\epsfig{figure=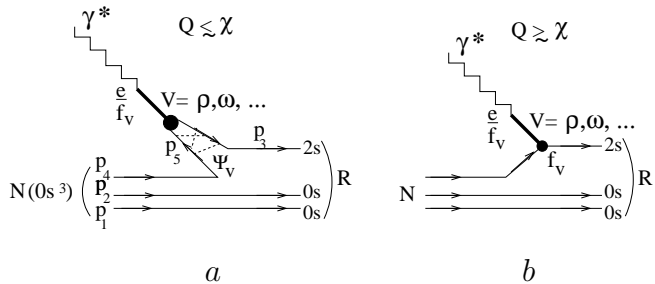,width=0.48\textwidth,clip}
\end{center}
\vspace{-1mm}\hspace{1cm}{\large $a$}\hspace{4cm}{\large $b$}
\caption{Diagrams of (a) ``soft'' (non-local) and (b) ``hard'' (local) 
coupling of vector mesons to the nucleon quark core.}
\label{f1}
\end{figure}

In such combined approaches two types of electromagnetic transition operators
are used, the operator designed for the soft $Q^2$ region and one for hard 
values of $Q^2$. However, in the transition amplitude they are summed for any
value of $Q^2$. For example, in the generalized $^3P_0+\,$EEM 
approach~\cite{cano98} the transition operator includes the sum of two 
vertices, schematically sketched in Figs.~\ref{f1}a and b. 

In the present paper we follow a more physical concept (see, e.g., 
Ref.~\cite{gross08} where the constituent quark and parton approaches to the
$\gamma qq$ vertex are discussed in the context of the nucleon electromagnetic
form factors). We can consider that the diagram  in Fig.~\ref{f1}a represents 
the unknown large-distance physics described by a phenomenological model
(the $^3P_0$ model in our case), which is adjusted to low-energy data 
(i.e. meson-nucleon coupling constants $\pi NN$, $\rho NN$, magnetic moments 
and decay widths). In the hard $Q^2$-region these contributions become less
important and an adequate description of the electromagnetic transition will 
be given by the diagram in Fig.~\ref{f1}b. In this case the unknown 
short-range physics is encoded by adjusted parameters of a parton model. 
In the region of moderate values of $Q^2$ 
(1.5$\,\lesssim Q^2\lesssim\,$4 GeV$^2$) 
we can consider the constituent quarks as partons and corresponding unknown 
short-range physics can be included in a few constituent quark parameters 
(such as quark form factors given by the intermediate vector mesons 
in the VMD model and scale parameters of quark configurations in the baryons). 
In this case it is not necessary to sum the contributions of the two diagrams 
in Fig.~\ref{f1}. Instead it would be desirable to use some mechanism for a 
smooth transition from one regime to the other. 

In our opinion such a mechanism can be described in general by a smooth 
transition from a typical hadron 
radius $b_V\approx\,$0.5 fm of the vector meson in the CQM to a point-like 
vector meson $b_V=\,$0 corresponding to the quark-parton picture sketched 
in Fig.~\ref{f1}b. Here we use the approximation 
$b_V(Q^2)=b_V(0)e^{-Q^2/\chi^2}$, where $\chi\simeq\, 1 - 2$ GeV 
corresponds to the lowest characteristic value of $Q^2$ where
the parton model phenomenology in deep inelastic $ep$ scattering sets in. 

Another important issue related to the Roper resonance is a possible combined
structure of this state which implies a virtual hadron-hadron component
(e.g. $\sigma N$ or/and $\pi\Delta$)~\cite{hanhart00} in addition to the 
radially excited
three-quark structure. Here we consider an admixture of the hadronic molecular 
state $N+\sigma$ in an effective description of such a component. We also 
consider to what degree such a combined structure for the Roper is compatible 
with the new high-quality data of JLab.

\section{Composite structure of the Roper resonance}\label{s2}

We consider the Roper resonance $(R)$ as a superposition  
of the radially excited three-quark configuration $3q^\ast$ and the
hadron molecule component $N+\sigma$ as: 
\eq 
|R\ra = \cos\theta |3q^\ast\ra + \sin\theta |N+\sigma\ra\,, 
\label{a32}
\en 
where $\theta$ is the mixing angle between the $3q^\ast$ and the hadronic 
component: $\cos^2\theta $ and $\sin^2\theta$ represent the 
probabilities to find a $3q^\ast$ and hadronic configuration, respectively. 
The parameter $\theta$ is adjusted to optimize the description of data on
the Roper resonance electroproduction. The limiting case of $\cos\theta = 1$ 
corresponds to the pure $3q^\ast$ interpretation of $R$, while the value 
$\cos\theta=0$ corresponds to the situation, where $R$ is a pure loose bound 
state of $N + \sigma$ (analogous to the deuteron -- bound state of proton and 
neutron). Note, in a first step we simplify the model by reducing it to two 
independent (decoupled) systems, $R_1=3q^\ast$ and $R_2=N+\sigma$, and do not 
consider the full coupled channel problem. Moreover, we consider the dynamics 
of the $R_1$ component in the framework of the nonrelativistic $^3P_0$ model 
(see, e.g. Refs.~\cite{gavella80,ackleh96}), while the dynamics of the $R_2$ 
component is considered in the framework of the hadronic molecular 
approach~\cite{HMM} which is manifestly Lorentz invariant. In future we intend 
to improve the description of the $R_1$ component by applying a relativistic 
quark model.  

First we briefly outline the basic notions of the $^3P_0$ mo\-del. 
The effective 
interaction term of the $^3P_0$ model~\cite{ackleh96,kalashnikova05} is set up 
as 
\eq 
H^{\rm eff}_{q}=g_q\int d^3x\bar{\psi_q}\psi_q\,, \quad\quad  
g_q=2m_q\gamma\,,
\label{a8}
\en 
where $\gamma$ is dimensionless constant. It can be considered as a static 
variant of the coupling $\gamma^\prime\bar q(x)q(x)S(x)$ 
where the external field $S(x)$ represents some scalar combination 
of gluon fields in the hadron. At low energy, where the dynamics is 
ruled by nonperturbative QCD, we pass to an effective description 
in terms of constituent quarks $\psi_q(x)$ and substitute a constant 
for the field $S(x)$. 

Apart from some drawbacks [see, e.g. Eqs.~(\ref{apr}) and 
(\ref{a7a}) in Appendix], the $^3P_0$ model~\cite{ackleh96,micu69,yaouanc73} 
is a good phenomenological method for the evaluation of hadron 
transitions~\cite{barnes97,capstick94,downum06,gutsche92} 
on the basis of the quark model starting from Eq.~(\ref{a8}) with 
a single strength parameter $\gamma$. The interaction term (\ref{a8}) gives 
rise to Feynman amplitudes for the $\bar qq$-pair creation (annihilation)
\eq 
& &(2\pi)^3\delta^{(3)}({\bf p}_4+{\bf p}_5)i{\cal M}^{\rm eff}_{fi} 
\nonumber\\
&=&
\langle q,\,{\bf p}_4,\mu_4|
\langle \bar q,\,{\bf p}_5,\mu_5|\,i\!\int d^3x 
{\cal L}^{\rm eff}_{q}(x)|0\rangle,
\label{a9}
\en 
which are used here for the calculation of meson-baryon couplings. The quark 
is labelled by its 3-momentum ${\bf p}_4$ 
and spin projection $\mu_4$ (for simplicity the isospin projection $t_4$ and 
the color are omitted), similarly for the antiquark. For the numbering of the 
quarks see Fig.~\ref{f1} (or Fig.~\ref{f6} in Appendix ~\ref{ap2}). 

The corresponding non-relativistic interaction term $V^{\rm eff}_q$ is defined 
as 
\eq 
\hspace*{-.2cm}
T^{\rm eff}_{fi}=\,_{nr}\!\langle q,\,{\bf p}_4,\mu_4|\,_{nr}\!
\langle \bar q,\,{\bf p}_5,\mu_5|
V^{\rm eff}_q|0\rangle \doteq \frac{1}{2m_q}{\cal M}^{\rm eff}_{fi},
\label{a11}
\en 
where a noncovariant normalization
\eq 
_{nr}\!\langle{\bf p},\mu|
{\bf p}^\prime,\mu^\prime\rangle\!_{nr}=
(2\pi)^3\delta^{(3)}({\bf p}-{\bf p}^\prime)\delta_{\mu,\mu^\prime}
\label{a11a}\
\en
is implied instead of the covariant one of Eq.~(\ref{a9}). 

Substitution of the non-relativistic reduction of the effective 
interaction (\ref{a8}) into Eqs. (\ref{a9}) and (\ref{a11}) 
leads to the expression
\eq 
V^{\rm eff}_q&\doteq&\frac{g_q}{2m_q}(-1)^{1-\mu_5-t_5}
\la {\scriptstyle\frac{1}{2}}-\!\mu_5|{\bm\sigma}
\!\cdot\!({\bf p}_4\!-\!{\bf p}_5)
|{\scriptstyle\frac{1}{2}}\mu_4\ra \nonumber\\
&\times&\la {\scriptstyle\frac{1}{2}}-\!t_5|{\scriptstyle\frac{1}{2}}t_4
\ra (2\pi)^3\delta^{(3)}({\bf p}_4+{\bf p}_5),
\label{a12}
\en 
which is the nonrelativistic analogue of the $\bar qq$ pair creation 
(annihilation) operator. 

The description of the hadronic $N+\sigma$ component of the Roper resonance
is based on the compositeness condition~\cite{Weinberg:1962hj,Efimov:1993ei}.   
This condition implies that the renormalization constant of the hadron 
wave function  is set equal to zero or that the hadron exists as a bound 
state of its constituents only. In the case of mixed states (as in the present
situation where the Roper is a superposition of the $3q^\ast$ and 
$N+\sigma$ components) the amplitude for the $N+\sigma$ component is defined 
by the parameter $\sin\theta$. The compositeness condition was originally  
applied to the study of the deuteron as a bound state of proton and 
neutron~\cite{Weinberg:1962hj}. Then it was extensively used 
in low--energy hadron phenomenology as the master equation for the 
treatment of mesons and baryons as bound states of light and heavy 
constituent quarks (see e.g. Refs.~\cite{Efimov:1993ei,Anikin:1995cf}).  
By constructing a phenomenological Lagrangian including the  
couplings of the bound state to its constituents and of the constituents  
to other particles in the possible decay channels we calculated hadronic-loop  
diagrams describing different decays of the molecular states  
(see details in~\cite{HMM}). 

\begin{figure}[hp]
\begin{center}
\epsfig{figure=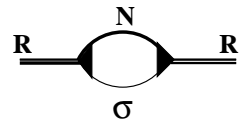,width=0.17\textwidth,clip}
\end{center}
\caption{The $N\sigma$ loop diagram contributing to the Roper mass operator.}  
\label{f2}
\end{figure}

In the present case the $R_2\to N+\sigma$ coupling is fixed from the 
compositeness condition 
\eq 
Z_{R} = 1-\Sigma^{\,\prime}_{N\sigma}(p)\vert_{\not p=m_R}=0,
\label{a33}
\en 
where $\Sigma_{N\sigma}(p)$ is the mass operator of the
$N\sigma$ bound state (Fig.~\ref{f2}), calculated with the use of the
phenomenological Lagrangian 
\eq 
{\cal L}_{R}^{\rm str}(x)
&=&g_{{\scriptscriptstyle R}\sigma {\scriptscriptstyle N}} 
\bar R(x) \int dy\Phi_R(y^2) \nonumber\\
&\times& N(x+w_{_{\sigma N}}y) \sigma(x-w_{_{N\sigma}}y) 
+{\rm H.c.}\,,
\label{a34}
\en 
where 
$w_{ij} = m_i/(m_i + m_j)$. 
Here $\Phi_R(y^2)$ is the correlation function describing the
distribution of $N\sigma$ inside $R$, which depends on 
the Jacobi coordinate $y$. Its Fourier transform used in the 
calculations has the form of a ``modified'' Gaussian, i.e. 
the Gaussian multiplied by a polynomial. In Euclidean space 
it may be written as
\eq
\tilde{\Phi}_R(-k_E^2)=\left(1-\lambda\frac{k_E^2}{\Lambda^2_M}\right)
\, \exp\left(-\frac{k_E^2}{\Lambda^2_M}\right)\,, 
\label{a35}
\en 
where $k_E$ is the Euclidean momentum. This present a kind of generalization 
of the nonrelativistic quark model wave function to the 4-dimensional case. 
But  the relativistic parameters $\lambda$ and 
$\Lambda_M$ should differ from the corresponding nonrelativistic 
ones. Here $\Lambda_M$ is the molecular size parameter and $\lambda$ is a free 
parameter which should be fixed by the orthogonality condition, i.e.  
$\la N| R \ra=\,$0. 


\section{Roper electroproduction}\label{s3}

\begin{figure}[hp]
\begin{center}
\epsfig{figure=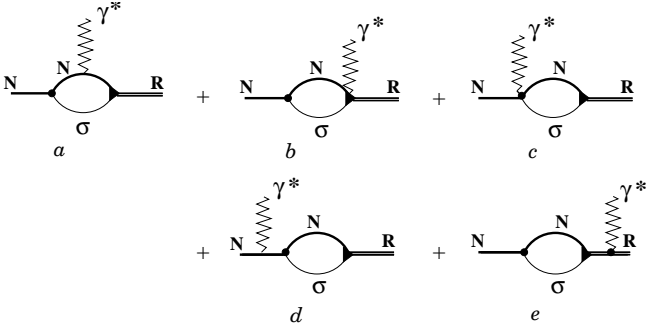,width=0.48\textwidth,clip}
\end{center}
\caption{$N\sigma$ hadronic-loop diagrams contributing to the 
Roper electroproduction: the triangle diagram (a), the bubble diagrams (b) 
and (c), the pole diagrams (d) and (e). 
} 
\label{f3}
\end{figure}

The diagrams which contribute to the Roper resonance electroproduction are
shown in Fig.~\ref{f1} (contribution of the $3q^\ast$ component) and 
Fig.~\ref{f3} (contribution of the hadronic $N\sigma$ component). In the
following we discuss the separate contributions of the structure components
of the Roper resonance.

\subsection{Contribution of the $3q^\ast$ component}

The contribution of the $3q^\ast$ component to the 
hadronic current of the Roper electroproduction is generally given as
\eq 
J^\mu=\la R|j^\mu_q|N\ra \doteq \la R, {\bf p'}, S_z', T_z'|j^\mu_q 
|N, {\bf p}, S_z, T_z\ra. 
\label{a36}
\en
The current $j^\mu_q$ is derived by starting from the vector meson absorption 
amplitudes described in the $^3P_0$ model
\eq
T^{q(\lambda)}_{V\!+\!N\to R}=3_{\,nr\!}\langle R,{\bf 0},S_z', T_z'|V^{eff}_q
|N,-{\bf q},S_z, T_z\rangle\nonumber\\
\times|V,{\bf q},\lambda_V,t_V\rangle_{nr}
\label{a36a}
\en
[see 
Appendix \ref{ap2} for details] and use of the vector meson dominance (VMD) 
mechanism in the photon-quark coupling: 
\eq 
e \, \epsilon^{(\lambda)}_\mu J^\mu 
=\frac{e}{2} \sum\limits_{V=\rho,\omega} 
\frac{{\cal M}^{q(\lambda)}_{V\!+\!N\to R}}{g_{\scriptscriptstyle VNN}} \, 
\frac{M_V^2}{Q^2+M_V^2}\,.
\label{a37}
\en 
The vector meson-nucleon coupling constant $g_{\scriptscriptstyle VNN}$ is 
calculated in the $^3P_0$ model [see Appendix \ref{ap2}] and we use
${\cal M}^{q(\lambda)}_{V\!+\!N\to R}
=2m_M\sqrt{2M_V}T^{q(\lambda)}_{V\!+\!N\to R}$ by taking into account 
a noncovariant normalization (\ref{a11a}) in Eq.~(\ref{a36a}).
$M_V$ is the vector meson mass approximated as 
$M_V=M_\rho\approx M_\omega$; ${\bf p}$, $S_z$, $T_z$ (${\bf p'}$, 
$S_z'$, $T_z'$) and ${\bf q}$, $\lambda_\rho$, $t_\rho$ are 
the 3-momentum, spin and isospin projections
of the nucleon (the Roper) and of the vector meson, respectively.  
For convenience we choose the photon momentum as $q^\mu=(q_0,0,0,|{\bf q}|)$. 

After substitution of the quark substructure (see Appendix \ref{ap1}) for
$|N\rangle$, $|R\rangle$ and $|V\rangle$ into Eqs.~(\ref{a36}) -- (\ref{a37}) 
and with a simple algebra (here we use $m_q=m_N/3$ for the quark mass), the 
current matrix element is deduced in the form 
\eq 
J^\mu\epsilon_\mu^{(\lambda)}&=&\frac{\sqrt{3}}{2}
\biggr[\frac{n(y_o)}{n(y)}\biggl]^{3/2}
e^{-\zeta(y){\bf q}^2b^2/6}\frac{M_V^2}{Q^2+M_V^2}\nonumber\\
&\times&\!\biggl\{\langle\frac{1+\tau_{z}}{2}\rangle_T \
\delta_{{\scriptstyle S}^\prime_z,{\scriptstyle S}_z} 
\biggl[\left(\epsilon_0^{(\lambda)}+
\frac{{\bf q}\!\cdot\!\bm{\epsilon}^{(\lambda)}}{2m_Nn(y)}\right)
\nonumber\\
&\times&p_2(y,{\bf q}^2)+
p_0(y)\frac{{\bf q}\!\cdot\!\bm{\epsilon}^{(\lambda)}}{2m_Nn(y)}\biggr]
\nonumber\\
&- &\!\!\langle\frac{1+5\tau_{z}}{2}\rangle_T \
\langle\frac{i[{\bm\sigma}\times{\bf q}]\!\cdot\!\epsilon^{(\lambda)}}{2m_N}
\rangle_S \ p_2(y,{\bf q}^2)\!\biggr\}\,. 
\label{a37a}
\en
We use the notation
\eq
\langle \cdots\rangle_S = 
\langle{\scriptstyle\frac{1}{2}}{\scriptstyle S}^\prime_z|\cdots 
|{\scriptstyle\frac{1}{2}}{\scriptstyle S}_z\rangle\,, \ \ 
\langle \cdots\rangle_T =  
\langle{\scriptstyle\frac{1}{2}}{\scriptstyle T}^\prime_z|\cdots 
|{\scriptstyle\frac{1}{2}}{\scriptstyle T}_z\rangle \nonumber 
\en
for the spin and isospin matrix elements, respectively, 
$\epsilon^{(\lambda)\mu}=\{\epsilon_0^{(\lambda)},\bm{\epsilon}^{(\lambda)}\}$ 
is the photon polarization vector and $n,\,\zeta,\,p_{0,2}$ are polynomials 
in $y=b_V/b$ and ${\bf q}^2$:
\eq 
n(y)=1+\frac{2}{3}y^2,\,\, \zeta(y)=\frac{1\!+\!\frac{5}{6}y^2}{n},\,\,
p_0(y)=\frac{4}{3}\frac{1\!+\!y^2}{n},\quad\nonumber\\
p_2(y,{\bf q}^2)=\frac{2}{3}\frac{y^2}{n}-
\left(\frac{1\!+\!y^2}{n}\right)^{\!\!2}\frac{{\bf q}^2b^2}{9}.
\qquad\qquad\qquad
\label{0}
\en 
The transverse ($\lambda=\,\pm$1) and longitudinal ($\lambda=$ 0) helicity 
amplitudes for electroproduction of the Roper resonance on the proton 
($T_z=1/2$) are defined by the matrix elements (\ref{a37a}) for $\lambda=$ +1 
and 0 respectively~\cite{capstick95,aznauryan07,romalho10}
\eq 
\hspace*{-.2cm}
A_{1/2}&=&\sqrt{\frac{2\pi\alpha}{q_{\scriptscriptstyle R}}}
\langle R,{\bf 0},+{\scriptstyle\frac{1}{2}}|\,
j^\mu_q\epsilon^{(+)}_\mu|
N,-{\bf q},-{\scriptstyle\frac{1}{2}}\ra \,,\nonumber\\
\hspace*{-.2cm}
S_{1/2}&=&\sqrt{\frac{2\pi\alpha}{q_{\scriptscriptstyle R}}}
\langle R,{\bf 0},+{\scriptstyle\frac{1}{2}}|
\,j^\mu_q\epsilon^{(0)}_\mu|
N,-{\bf q},+{\scriptstyle\frac{1}{2}}\rangle\frac{|{\bf q}|}{Q}
\label{a41}
\en 
where $\alpha = 1/137$ is the fine-structure constant. We introduce 
\eq 
q_R = \frac{m_R^2-m_N^2}{2m_R} 
\en 
for the threshold value of the photon 3-momentum for Roper electroproduction.

In the rest frame of the Roper resonance (the c.m. frame of $\gamma^*N$ 
collision) the absolute value of the transfered three-momentum ${\bf q}$ in 
Eqs.~(\ref{a41}) is defined by 
\begin{equation}
{\bf q}^2=Q^2+\left(\frac{Q^2+m_N^2-m_R^2}{2m_R}\right)^2.
\label{a41a}
\end{equation}
Note that in the region of 0.5$\lesssim Q^2\lesssim\,$1.5 GeV$^2$ the c.m. 
frame is very close to the Breit frame, i.e. $Q^2\approx {\bf q}^2$, which is 
very convenient for comparison of our ${\bf q}^2$-dependent results with the
relativistic $Q^2$-dependent ones (substitution of ${\bf q}^2\to Q^2$ does not
really change our results if one considers the region 
0.5$\lesssim Q^2\lesssim\,$1.5 GeV$^2$). 

We have several remarks regarding current conservation connected to the 
gauge symmetry of theory. 
The current conservation condition $q_\mu J^\mu = 0$ for the matrix 
elements (\ref{a36}) is not automatically satisfied for the VMD amplitudes. 
To provide $q_\mu J^\mu = 0$ for a transition current in the VMD amplitudes 
one needs the conservation of the neutral vector meson currents 
$\partial^\mu J_\mu^V=0$ \cite{Kroll_Lee_Zumino}. In our model these currents 
$J_\mu^V$ are expressed via the amplitudes (\ref{a36a}) for which the 
relation $J^0=\frac{|{\bf q}|}{q_0}J^3$ is not exactly satisfied. 

One can try to construct the electromagnetic current using the transverse 
projector $g_\perp^{\mu\nu}=g^{\mu\nu}+\frac{q^{\mu}q^{\nu}}{Q^2}$. 
This projector does not change the $A_{1/2}$ amplitude, but could lead to some 
corrections for the components $J^0$ and $J^3$. However, the expression for 
$S_{1/2}$ in Eq.~(\ref{a41}) is invariant with respect to such corrections 
because of the contraction of the current matrix elements with the 
longitudinal polarization vector
\eq
\epsilon^{(0)\mu}=\bigl\{\frac{|{\bf q}|}{Q},0,0,\frac{{q_0}}{Q}\bigr\}.
\label{a41b}
\en
Some problem appears in a small region near 
the photon point with $Q^2=0$ where 
the last factor for $S_{1/2}$ in expression (\ref{a41}) 
shows singular behavior. In this region we use the following trick. 
We start from the exact equality 
\eq
J^0=J^3 \quad\mbox{at}\,\,Q^2=0
\label{ward}
\en 
which follows from the Ward identity at the photon point where 
$q^0=|{\bf q}|$. Note that at $Q^2=0$ we really have $J^0\approx J^3$ if we 
use realistic parameters of the constituent quark model (CQM) for the wave 
functions of baryons and mesons. Thus it is not difficult to transform the 
approximate equality $J^0\approx J^3$ to the exact one of Eq.~(\ref{ward}) by 
slightly varying one of the free parameters of the CQM (e.g. the radius $b_R$ 
of the quark core of the Roper resonance which is not strictly fixed 
otherwise). The constraint (\ref{ward}) imposed on the parameters of the quark 
wave functions in the $^3P_0$ amplitudes only stabilizes the behavior of 
$S_{1/2}$ near the photon point $Q^2\lesssim\,$0.2 - 0.3 GeV$^2$ and does not 
give pronounced effects for $S_{1/2}$ in the remaining region for $Q^2$, where 
$\frac{|{\bf q}|}{Q}\approx\,$1.
 
Our results for the helicity amplitudes are: 
\eq
A_{1/2}&=&-\sqrt{\frac{2\pi\alpha}{q_{\scriptscriptstyle R}}}
\frac{\sqrt{3}}{2}\mu_p\langle\sigma_+\rangle
\biggr[\frac{y_{\scriptscriptstyle R}n(y_0)}
{N(y,y_{\scriptscriptstyle R})}\biggl]^{3/2} 
\frac{M_V^2}{Q^2+M_V^2}\nonumber\\
&\times& e^{-\tilde\zeta(y,y_{\scriptscriptstyle R})
\frac{{\bf q}^2b^2}{6}} \, 
\frac{|{\bf q}|}{2m_N}P_2(y,y_{\scriptscriptstyle R},{\bf q}^2)
\quad
\label{a42}
\en 
and 
\eq 
S_{1/2}&=&-\sqrt{\frac{2\pi\alpha}{q_R}}\frac{\sqrt{3}}{2}
\biggr[\!\frac{y_{\scriptscriptstyle R}n(y_0)}
{N(y,y_{\scriptscriptstyle R})}\biggl]^{3/2}\!\!\!\!\!
\frac{M_V^2}{Q^2+M_V^2}
e^{-\tilde\zeta(y,y_{\scriptscriptstyle R})\frac{{\bf q}^2b^2}{6}}
\nonumber\\
&\times&\frac{{\bf q}^2}{Q^2}\biggl\{ 
\bigl[1+\frac{q_0(\frac{3}{2}y^2_{\scriptscriptstyle R}\!-\!\frac{1}{2})}
{2m_NN(y,y_{\scriptscriptstyle R})}\bigr]
P_2(y,y_{\scriptscriptstyle R},{\bf q}^2)
\nonumber\\
&+&\frac{q_0}{2m_NN(y,y_{\scriptscriptstyle R})}
P_0(y,y_{\scriptscriptstyle R})\!\!\biggr\}\quad
\label{a45}
\en 
where $y \equiv y(Q^2) = y_0 \exp(-Q^2/\chi^2)$. We also take into account 
a possible difference of the $R$-resonance radius $b_{\scriptscriptstyle R}$ 
and the one of the nucleon, which is $b$, by introducing the ratio 
$y_{\scriptscriptstyle R}=b_{\scriptscriptstyle R}/b$ which does not depend on 
$Q^2$. As a result the polynomials (\ref{0}) become
$y_{\scriptscriptstyle R}$-dependent ones now denoted as 
$N,\,P_{0,2},\,\tilde\zeta$ [see Eqs.~(\ref{a28}) -- (\ref{a28a}) in the 
Appendix] and only for $y_{\scriptscriptstyle R}=1$ they are identical with
$n,p_{0,2},\zeta$:
\eq
&&n(y)=N(y,y_{\scriptscriptstyle R}=1),\quad 
p_2(y,{\bf q}^2)=P_2(y,y_{\scriptscriptstyle R}=1,{\bf q}^2),\nonumber\\
&&p_0(y)=P_0(y,y_{\scriptscriptstyle R}=1),\quad
\zeta(y)=\tilde\zeta(y,y_{\scriptscriptstyle R}=1).
\label{a45a}
\en 
\begin{figure}[hp]
\begin{center}
\epsfig{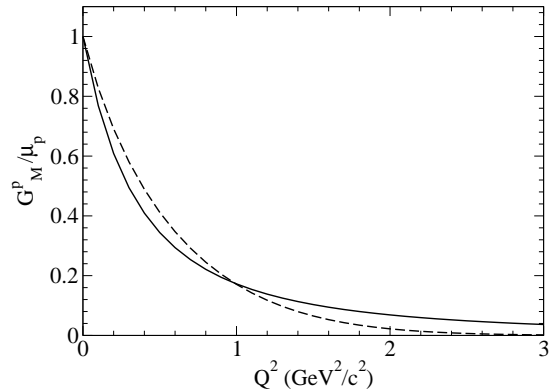}
\end{center}
\caption{Normalized magnetic form factor of the proton $G^p_M/\mu_p$ in the 
modified $^3P_0$ model with a $Q^2$-dependent vector meson radius and the VMD 
approach to the $qq\gamma$ interaction (dashed line). Here we use the same 
set of parameters as in the $N\to R$ vertex of Fig.~\ref{f5}a.
For comparison, the dipole approximation is also shown (solid line).}
\label{f4}
\end{figure}

The vector meson contribution to the amplitude at high $Q^2$ should also 
contain contributions of vector mesons of higher mass 
$M_V\gtrsim$ 2$M_\rho$. Following the work~\cite{gross08} we use the 
approximation
\eq 
&&\frac{M_V^2}{Q^2+M_V^2}=x\frac{M_\rho^2}{Q^2+M_\rho^2}+
(1-x)\frac{4M_\rho^2}{Q^2+4M_\rho^2}\,, \nonumber\\
&&x=0.7\,, 
\label{a43}
\en 
which we have checked in the description of the nucleon magnetic form factor.

Note the matrix element (\ref{a36}) for the diagonal transition 
$N+\gamma^*\to N$ has the same form as Eq.~(\ref{a37a}) excluding 
the algebraic factor $\frac{\sqrt{3}}{2}$ and the polynomial $p_2$ 
which should be changed to 1. 
In the static limit $|{\bf q}|,q_0\to\,$0 this expression defines
the charge and the magnetic moment of the nucleon (for $m_q=m_N/3$)
\eq
\hat e=e\frac{I+\tau_z}{2},\quad 
\hat\mu=\mu_N\,\frac{I+5\tau_z}{2},\quad \mu_N=\frac{e}{2m_N}.
\label{a39}
\en
The values of $\mu_p$ and $\mu_n$ are reproduced with an accuracy of about 
10\%. Moreover, at low and moderate values of $Q^2$ this amplitude describes 
the nucleon magnetic form factor $G_M$ with a reasonable accuracy (see 
Fig.~\ref{f4}). Such an accuracy is sufficient (at least in the region 
$Q^2\lesssim\,$1 - 1.5 GeV$^2$) for the present calculation of the Roper 
electroproduction amplitudes. 

For the non-diagonal process $N+\gamma^*\to R$ the matrix element (\ref{a37a}) 
defines `the transition magnetic moment` in the limit 
$|{\bf q}|,q_0\to q_{\scriptscriptstyle R}$ (i.e. at the photon point):
\eq
\hat\mu_{N\to R}=-\frac{e}{2m_N}\,\frac{(I+5\tau_z)}{2}\,
\frac{\sqrt{3}}{2}exp[-\zeta(y_0){\bf q}^2_{\scriptscriptstyle R}b^2/6]
\nonumber\\
\times\biggl[\frac{2y_0^2/3}{1\!+\!2y_0^2/3}
-\left(\!\frac{1\!+\!y_0^2}{1\!+\!2y_0^2/3}\!\right)^{\!2}
\frac{{\bf q}^2_{\scriptscriptstyle R}b^2}{9}\biggr].
\label{a40}
\en
The quantity $\hat\mu_{N\to R}$ gives the value 
(apart from a kinematical factor
$\langle\sigma_+\rangle\sqrt{q_{\scriptscriptstyle R}/2}$) of
the transverse helicity amplitude $A_{1/2}$ at the photon point. 
The first term in the square brackets of the r.h.s. of Eq.~(\ref{a40})
\eq
Z_V=\frac{2y_0^2/3}{1\!+\!2y_0^2/3},\quad y_0=b_{\scriptscriptstyle V}/b
\label{a40a}
\en
(or the first term of the polynomial $p_2$ in Eq.~(\ref{0}))
is present because of the nonlocality of the $Vqq$ interaction 
defined by Eq.~(\ref{a36a}).
There the operator $V_q^{eff}$ leads to an insertion of the inner $\bar qq$ 
wave function of the vector meson into the $Vqq$ vertex. 

The size of the nonlocal region is defined by the spatial scale of the meson 
wave function. For a point-like vector meson ($b_{\scriptscriptstyle V}=\,$0) the 
value of $Z_V$ reduces to zero, and the matrix element for the transition 
$N+\gamma^*_{\scriptscriptstyle T}\to R$ reduces to the matrix 
element of the elementary-emission model (EEM) with a local $Vqq$ vertex.
The EEM matrix element vanishes in the limit $|{\bf q}|\to\,$0, as it 
should because of the orthogonality of the spatial parts of the wave functions 
of $N$ and $R$. 

Such behavior of the $A_{1/2}$ amplitude near the photon point $Q^2=\,$0
is characteristic of all the models which start from local $\gamma qq$ or 
$Vqq$ vertices at high $Q^2$ and continue to use such interaction in the `soft`
region of small $Q^2\lesssim 6/b_{\scriptscriptstyle V}^2$, where the e.-m. 
interaction is modified by the inner structure of vector mesons as  
$\bar qq$ bound states.

As a result, in models with a local operator for the $\gamma qq$ (or  $Vqq$)
interaction (see, e.g. the relativistic 
models~\cite{weber90,capstick95,cardarelli97,aznauryan07,romalho10}) the 
transverse helicity amplitude $A_{1/2}$ vanishes in the limit  $Q^2\to\,$0
[or it approaches a small value which is defined by the second term 
in the last line of Eq.~(\ref{a40}) modified by relativistic corrections]. 

The importance of the nonlocality of the $Vqq$ interaction in the description 
of Roper electroproduction near the photon point was first noted by the authors 
of the $^3P_0$ model~\cite{gavella80}. In Ref.~\cite{cano98} this nonlocal 
$^3P_0$ interaction was used for the calculation of the helicity amplitudes
on the basis of a dynamical quark model of baryons. Unfortunately, the authors 
of~\cite{cano98} have only used a trivial sum of $^3P_0$ and EEM interaction 
terms (a `generalized EEM`). With this ansatz they describe both the low- and 
high-$Q^2$ amplitudes with a common mechanism, and the same quark dynamics 
was used for both the nucleon and the Roper resonance.

Now it becomes evident that intermediate meson-baryon states (`hadron loops`)
can play a considerable role in the quark dynamics of excited baryons, and 
such meson-baryon states should be taken into account (see, e.g., 
Ref.~\cite{BRAG07,barnes08}). Since the resonance pole of the 
Roper~\cite{pdg10} 1365 - i95 MeV is rather close to the $N+\sigma$ threshold 
the intermediate $N+\sigma$ configuration will play a more important role in 
the inner dynamics of the Roper as compared for example to the case of the 
nucleon.

In our opinion, a first step in the study of the nontrivial inner structure of 
the Roper resonance could be an evaluation on the basis of the recent CLAS 
data~\cite{aznauryan09}, where a nonvanishing probability for a possible 
$N+\sigma$ component of the Roper is compatible with the data..

\subsection{Contribution of the hadronic $N + \sigma$ component}

The hadronic $N\sigma$ loop diagrams contributing to the Roper 
electroexcitation are shown in Fig.~\ref{f3}. The $RN\sigma$ vertex is defined
by the nonlocal Lagrangian ${\cal L}_{R}$ of Eq.~(\ref{a34}).
For the $NN\sigma$ vertex we use a similar nonlocal Lagrangian with the
correlation function $\Phi_N(y^2)$ 
\eq
\label{a34nucleon}
{\cal L}_{N}&=&g_{{\scriptscriptstyle NN}\sigma}\, \sigma(x) \, 
\int dy\,\Phi_N(y^2)  \nonumber\\
&\times& \bar{N}(x+y/2)N(x-y/2),
\en
where $g_{NN\sigma}$ is the $NN\sigma$ coupling constant, 
$\tilde{\Phi}_N(-k_E^2)=\exp{\left(-\frac{k_E^2}{\Lambda_N^2}\right)}$ 
is the Fourier transform of $\Phi_N(y^2)$ in Euclidean space 
with $\Lambda_N =\,$0.7 -- 1 GeV.

The electromagnetic interaction Lagrangian contains two pieces  
\eq 
{\cal L}_{\rm int}^{\rm em} = 
{\cal L}^{\rm em (1)}_{\rm int} + {\cal L}^{\rm em (2)}_{\rm int}
\en 
which are generated after the inclusion of photons. 
The first term ${\cal L}^{\rm em (1)}_{\rm int}$ is standard and is obtained by
minimal substitution in the free Lagrangian 
of the proton and charged Roper resonance: 
\eq
\label{photon}
\partial^\mu  B \to (\partial^\mu - ie_B A^\mu) B\,,
\en 
where $B$ stands for $ p, R^+$ and 
$e_B$ is the electric charge of the field $B$.  
The interaction Lagrangian ${\cal L}^{\rm em (1)}_{\rm int}$ reads 
\eq
\label{L_em_1}
{\cal L}^{\rm em (1)}_{\rm int}(x) =
  e_B \bar B(x) \!\not\!\! A \, B(x). 
\en
The second electromagnetic interaction term ${\cal L}^{\rm em (2)}_{\rm int}$
is generated when the nonlocal Lagrangians~(\ref{a34}) and (\ref{a34nucleon})
are gauged. The gauging proceeds in a way suggested and extensively used
in Refs.~\cite{Anikin:1995cf,Mandelstam:1962mi,Terning:1991yt}.
In order to guarantee local $U(1)$ gauge invariance of the strong interaction
Lagrangian one multiplies each charged field in~(\ref{a34}) and 
(\ref{a34nucleon}) with a gauge field exponential $e^{-ie_B I(y,x,P)}$. 
The exponent contains the term
\eq\label{path}
I(y,x,P) = \int\limits_x^{y} dz_\mu A^\mu(z)\,, 
\en 
where $P$ is the path of integration from $x$ to $y$. 
Then we obtain
\eq\label{gauging1}
{\cal L}_{R}^{\rm str + em(2)}(x)  
&=&g_{{\scriptscriptstyle R}\sigma {\scriptscriptstyle N}} 
\bar R^0(x) \int dy\Phi_R(y^2) \nonumber\\
&\times& n(x+w_{_{\sigma N}}y) \sigma(x-w_{_{N\sigma}}y) \nonumber\\
&+&g_{{\scriptscriptstyle R}\sigma {\scriptscriptstyle N}} 
\bar R^+(x) \int dy\Phi_R(y^2) \nonumber\\
&\times& e^{-ie_{p} I(x+w_{_{\sigma N}}y,x,P)}
p(x+w_{_{\sigma N}}y) \nonumber\\ 
&\times&\sigma(x-w_{_{N\sigma}}y) \, + \, {\rm H.c.}  
\en 
and 
\eq\label{gauging2}
{\cal L}_{N}^{\rm str + em(2)}(x) &=& 
g_{{\scriptscriptstyle NN}\sigma}\,
\sigma(x) \, \int dy\,\Phi_N(y^2)  \nonumber\\
&\times& \biggl(\bar{n}(x+\frac{y}{2})n(x-\frac{y}{2}) \\
&+& \bar{p}(x+\frac{y}{2})
e^{-ie_{p} I(x-\frac{y}{2},x+\frac{y}{2},P)}
p(x-\frac{y}{2}) \biggr).  \nonumber
\en 
An expansion of the gauge exponential up to terms linear in $A^\mu$
leads to ${\cal L}^{\rm em (2)}_{\rm int}$.

The full Lagrangian consistently generates the required matrix element of the 
electroexcitation amplitude which is linked to coming the hadronic molecular 
component of the Roper. 
Because of gauge invariance the electromagnetic vertex function 
$\Lambda_{\mu}(p,p')$ is orthogonal to the photon momentum 
$q^{\mu}\Lambda_{\mu}(p,p')=0$. As a result, the vertex function 
$\Lambda_\mu(p,p')$ is given by the sum of the gauge-invariant pieces of the
triangle $(\Delta)$, the bubble (bub) and the pole (pol) diagrams, 
while the non gauge-invariant parts of these diagrams cancel in the sum: 
\eq
\Lambda_{\mu}(p,p') &=&
\Lambda_{\mu, \, \Delta}^{\perp}(p,p') +
\Lambda_{\mu, \, {\rm bub}}^{\perp}(p,p') \nonumber\\
&+& 
\Lambda_{\mu, \, {\rm pol}}^{\perp}(p,p'). 
\en 
The contribution of each diagram can be split into a gauge invariant piece and 
a reminder term, which is not gauge invariant, by introducing the 
decomposition 
\eq\label{split}
\gamma_\mu \, = \, \gamma_\mu^{\perp} \, + \,
q_\mu \, \frac{\not\! q}{q^2}\,,  \hspace*{1cm}
p_{i\, \mu} \, = \, p_{i\, \mu}^{\perp} \, + \, q_\mu \, 
\frac{p_{i} q}{q^2}\,, 
\hspace*{1cm}
\en
with $\gamma_\mu^\perp \,  q^\mu \, = \, 0$, 
$p_{i\, \mu}^\perp \, q^\mu=0$, where $p_{i}$ is  $p$ or $p'$. 
The vertex function $\Lambda_\mu^\perp(p_{1},p_{2})$ can then be 
expressed in terms of $\gamma_\mu^\perp$ 
and $p_{i\,\mu}^\perp$. 

In the case of the triangle diagram of Fig.~\ref{f3}a we include the 
$q^2$-dependence of the photon-nucleon vertices in correspondence with data. 
Taking into account the nucleon structure the $e_p \bar p \gamma^\mu_\perp p$ 
vertex is modified as 
\eq
\label{NNGamma_vertex}
\bar{N}\left(\gamma_{\perp}^\mu\,F_1^N(q^2)\,+
\,i\frac{\sigma^{\mu\nu}q_{\nu}}{2m_N}\,
F_2^N(q^2)\right) N,
\en
where $F_1^N(q^2)$ and $F_2^N(q^2)$ are the Dirac and Pauli form factors, 
which are normalized as $F_1^N(0)=e_N$ (nucleon electric charge) and 
$F_2^N(0) = \kappa_N$ (nucleon anomalous magnetic moment). 
The form factors  
$F_{1,2}^N$ are expressed through the electric and magnetic Sachs form factors 
$G_E^N$, $G_M^N$ of the nucleon as 
$F_1^N=(G_E^N+\tau\,G_M^N)/(1+\tau)$, 
$F_2^N=(G_M^N-G_E^N)/(1+\tau)$, $\tau=-q^2/4m_N^2$. 
For the Sachs form factors we use the Kelly parametrization~\cite{Kelly}: 
\eq
\label{NNGamma_Kelly}
G(\tau)\propto \frac{\sum_{k=1}^n\,a_k\,\tau^k}{1+\sum_{k=0}^{n+2}
\,b_k\,\tau^k}.
\en

\begin{figure}[hp]
\begin{center}
\epsfig{figure=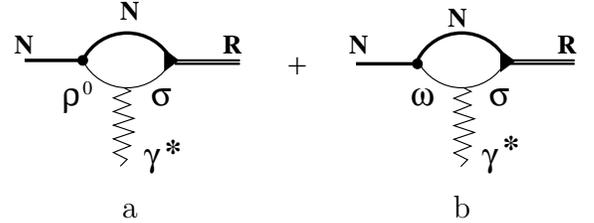,width=0.42\textwidth,clip}

\vspace{2mm}

{\large\mbox{a}\hspace{42mm}\mbox{b}}
\end{center}
\caption{$\sigma\gamma^\ast V$ ($V=\rho^0$, $\omega$) processes in the 
electroexcitation of the $N\sigma$ bound state.}
\label{f3a}
\end{figure}

Two additional contributing diagrams to the electroproduction of the Roper
resonance are shown in Fig.~\ref{f3a}. 
The amplitudes of the $\sigma\gamma^\ast V$ ($V=\rho^0$, $\omega$) 
transition are written in the form
\begin{equation}
\label{VSigmaGamma_amp}
\frac{e\,g_{\sigma\gamma V}}{M_V}
\left(g^{\mu\nu}\,q\cdot k-k^\mu\,q^\nu\right) \,, 
\end{equation} 
where $k$ is the vector meson momentum.  
The values for the coupling constants $g_{\sigma\gamma V}$ 
are estimated in the branch ratio model~\cite{Achasov1} with 
$g_{\sigma\gamma \rho^0}\simeq 0.25$, $g_{\sigma\gamma \omega}\simeq 0.05$.

The contributions of the amplitudes of Fig.~\ref{f3a} were estimated using the 
local limit for the $NN\rho$ and $NN\omega$ vertices. We found a very small 
contribution compared to the diagrams of Fig.~\ref{f3}. Both $\sigma\gamma V$ 
diagrams are explicitly transverse under contraction with the photon momentum 
$q_\mu$.

Finally, the helicity amplitudes for the electromagnetic excitation are 
defined like in 
(\ref{a41})
\begin{eqnarray}
\label{helicity_amp}
A_{1/2}&=&\sqrt{\frac{2\pi\alpha}{q_{\scriptscriptstyle R}}}\left\langle\,R,
\frac{1}{2}\,\right\vert\,J_{+}\,\left\vert\,N,-\frac{1}{2}\,\right\rangle\,
\xi \nonumber\\ 
S_{1/2}&=&\sqrt{\frac{2\pi\alpha}{q_{\scriptscriptstyle R}}}\left\langle\,R,
\frac{1}{2}\,\right\vert\,J_{0}\,\left\vert\,N,\frac{1}{2}\,\right\rangle\,
\xi ,\\ 
J_{+}&=&-\frac{1}{\sqrt{2}}(J_x+i\,J_y), \nonumber
\end{eqnarray}
where $J^{\mu}$ is the electromagnetic transition current defined by the 
diagrams of Fig.~\ref{f3}. The helicity amplitudes (\ref{helicity_amp}) are 
defined up to a phase $\xi$. The amplitudes are written in the c.m. frame of 
the nucleon and the photon, i.e. in the Roper-resonance rest frame. 
The 4-spinors present in $\vert\,R\,\rangle$, $\vert\,N\,\rangle$ are 
normalized as $\bar{R}R=\frac{m_R}{\varepsilon_R}$, 
$\bar{N}N=\frac{m_N}{\varepsilon_N}$.

\section{Results and comparison with data}\label{s4}
\subsection{Parameter fitting}
In the calculation the helicity amplitudes $A_{1/2}$ and $S_{1/2}$ we use two 
variants for the free parameters, denoted as (a) and (b), both typical for the
CQM. They were only fitted to the $A_{1/2}$ JLab 
data~\cite{aznauryan09,mokeev09} without any additional adjustment to the 
$S_{1/2}$ data [we only take into account the condition (\ref{ward})]. One of 
the parameter sets gives the best description of the data in the `soft` 
region with 0$\,\lesssim Q^2\lesssim\,$1 GeV$^2$ and the other one the optimal
description in the whole measured interval including the `hard` region of
$Q^2\gtrsim\,$1.5 -- 2 GeV$^2/c^2$. 

Note, we do not pretend that the non-relativistic model for the quark 
configurations is able to describe data for the whole `hard` region. 
We only study the compatibility of our predictions with the behavior of the 
data in the transition region between the `soft` and the `hard` regimes 
(a relativistic generalization of the model could be the next step to start
from a `hard` variant which gives a realistic description of the data at low 
and moderate values of $Q^2$). 

Our parameters are grouped into two sets: one set of parameters is related to 
the $3q$ components of the baryons and the other set is connected with the
$N\sigma$ molecular component. 
One of the quark model parameters is fixed by the strong constraints
following from the Ward identity [parameter $b_R$, see Eq.~(\ref{ward})].
The additional free quark parameters $b,\,b_{\scriptscriptstyle V}$ and $\chi$
are adjusted to optimize the description of the proton magnetic form
factor in the considered region of $Q^2$, including the intermediate values
0.5$\,\lesssim Q^2\lesssim\,$1.5 GeV$^2/c^2$, and of the above mentioned subset of
data on the helicity amplitude $A_{1/2}$.
Two of these fitted parameters, $b$ and $b_{\scriptscriptstyle V}$,
should in addition have values which are typical for the quark core radii of
the nucleon and a vector meson with
$b\approx b_{\scriptscriptstyle V}\approx\,$0.5 fm.
The third fitted parameter $\chi$ should not be smaller than 
the characteristic scale $\sim\,$(1 -- 1.5)$m_N$ associated with short-range 
effects in $eN$ scattering.
These additional constraints on the parameters $b$, $b_{\scriptscriptstyle V}$ and 
$\chi$ sufficiently limit the range of allowed values. 
Finally we arrive at the following two optimal sets of quark 
component parameters:

(a) a `hard` variant
\eq 
& & b=0.48\,\mbox{fm}\,, \  
    y_0=\frac{b_{\scriptscriptstyle V}}{b}=0.9\,, \ 
\chi^2=1.5\,m_{\scriptscriptstyle N}^2 \nonumber\\
& &b_{\scriptscriptstyle R}=0.9444\,b 
\label{a}
\en
adjusted to the data of $A_{1/2}$ with taking into account the `hard` region of
$Q^2\gtrsim\,$1.5 -- 2 GeV$^2/c^2$
and

(b) a `soft` variant
\eq 
& & b=0.54\,\mbox{fm}\,, \  
    y_0=\frac{b_{\scriptscriptstyle V}}{b}=0.81\,, \ 
\chi^2=4\,m_{\scriptscriptstyle N}^2 \nonumber\\
& &b_{\scriptscriptstyle R}=0.8824\,b\,,
\label{b}
\en
fitted to the $A_{1/2}$ data with  0$\,\lesssim Q^2\lesssim\,$1 GeV$^2/c^2$.  

The set of parameters related to the molecular component includes the mixing 
parameter $\theta$, the scale parameters $\Lambda_M$, $\Lambda_N$ and the 
parameter $\lambda$ entering in the vertex function of the Roper. Further 
parameters linked to the $\sigma$ are the mass $M_\sigma$, the width 
$\Gamma_\sigma$ and the strong coupling constant $g_{\sigma NN}$. The parameters 
$\Lambda_M\approx\Lambda_N\approx\,$1~GeV are approximately taken at the
scale set by the light baryons. The parameter $\lambda$ is fixed through the 
orthogonality condition $\langle R|N\rangle=\,0$ 
(finally fitted at $\lambda=2.45$). For the $\sigma$ resonance 
we take values which are reasonable~\cite{pdg10}  
(a wide range of values is 
given by $M_\sigma = (0.4 - 1.2)$~GeV, $\Gamma_\sigma = (0.5 - 1)$~GeV and
$g_{\sigma\scriptscriptstyle NN}\approx\,$5 - 10). 
Some fine-tuning of these parameters to the complete range of data on 
$A_{1/2}$ results in the following set of molecular parameters:
\eq 
& &\ \Lambda_M=1\,\mbox{GeV}\,, \
\Lambda_N=0.8\,\mbox{GeV}\,,\nonumber\\ 
& &M_\sigma=0.5 \pm 0.05 \,\mbox{GeV}\,, \ \ 
\Gamma_\sigma=0.75 \pm 0.25 \,\mbox{GeV}\,, \nonumber\\
& &g_{\sigma{\scriptscriptstyle NN}}=5\,. 
\label{mol}
\en 
The mixing parameter $\theta$ is fixed in the low energy region 
(0$\lesssim Q^2\lesssim$1 GeV$^2/c^2$) of $A_{1/2}$, where 
the molecular component is optimized to reproduce the differnece between the
$3q$ contribution and the JLab data. We obtain
$\sin{\theta}=\,$0.6 and 0.7 for sets (a) and (b) respectively.
The complete results for the parameters should be considered preliminary and 
be tested seriously in further applications.

It is important to remark that in the evaluation of the helicity amplitudes 
we use the free $\sigma$ meson propagator (as some kind of approximation), 
while in case of the strong Roper decay $R \to N + 2\pi$ we had to use the 
Breit-Wigner $\sigma$-meson propagator. 
The sensitivity of the results to a variation of the $\sigma$ meson mass 
from 0.45 to 0.55 GeV gives a variation of the helicity amplitudes up to 
$10\%$. The sensitivity of the strong decay $\Gamma(R \to N + 2\pi)$ 
to a variation of $\Gamma_\sigma$ is discussed in Sec.IVC. 
In fact, more precise data on $\Gamma(R \to N + 2\pi)$ can give a new,  
additional constraint on $\Gamma_\sigma$. 

\subsection{Helicity amplitudes}
\begin{widetext}

\begin{figure}[hp]
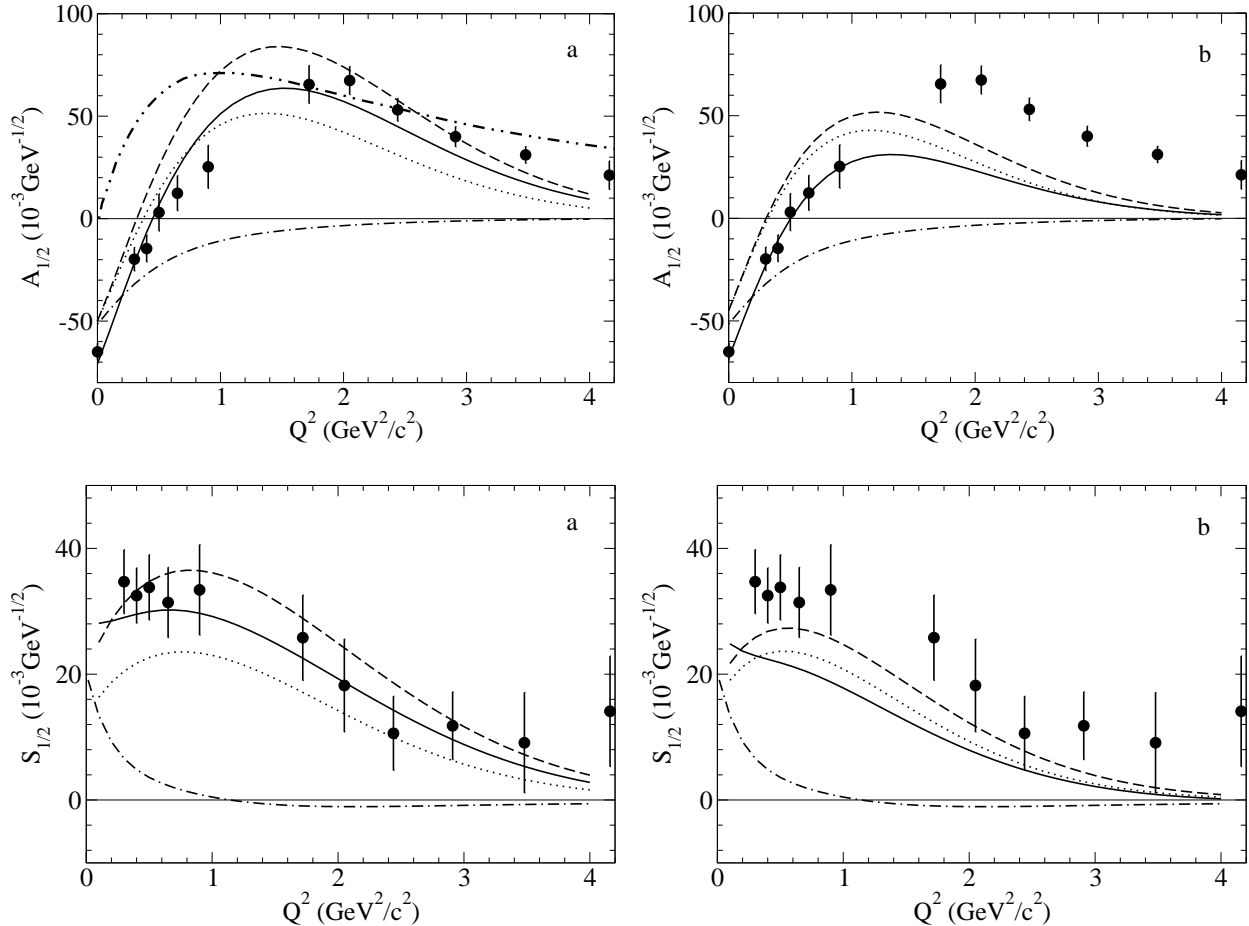





\begin{center}
\mbox{
{\epsfig{figure=a12aG.eps,width=0.45\textwidth,clip}}\quad
{\epsfig{figure=a12b.eps,width=0.45\textwidth,clip}}
}

\vspace*{0.5cm}

\mbox{
{\epsfig{figure=s12a.eps,width=0.45\textwidth,clip}}\quad
{\epsfig{figure=s12b.eps,width=0.45\textwidth,clip}}
}
\caption{Helicity amplitudes $A_{1/2}$ (top panels) and $S_{1/2}$ (bottom) for 
two variants of the model parameters, `hard` (a, left panels) and `soft` 
(b, right panels), in comparison to JLab data~\cite{aznauryan09,mokeev09}.
Dotted curves --- the quark core excitation amplitudes 
$|3q\rangle+\gamma^*\to|3q^*\rangle$ calculated in the framework of the 
standard `$^3P_0+$ VMD' model with a fixed vector meson radius 
$b_{\scriptscriptstyle V}=y_{\scriptscriptstyle 0}b$. Dashed curves --- the same
amplitudes calculated in a modified `$^3P_0+$ VMD' model with a $Q^2$-dependent 
scale parameter $y=y_0e^{-Q^2/\chi^2}$ for the vector meson radius 
$b_{\scriptscriptstyle V}=yb$.
Dashed-dotted curves --- helicity amplitudes for the electroexcitation of the
hadron molecule $N+\sigma$ . Solid curves --- the full calculation of 
$A_{1/2}$ and $S_{1/2}$ in terms of a combined structure  
$R=\cos\theta|3q^*\rangle+\sin\theta|N+\sigma\rangle$.
For comparison, the valence quark contribution to $A_{1/2}$ calculated in 
Ref.~\cite{romalho10} on the basis of a covariant spectator model is also 
shown (the dashed-double-dotted curve in the left top panel).}
\label{f5}
\end{center}
\end{figure}

\end{widetext}

The calculated helicity amplitudes $A_{1/2}$ and $S_{1/2}$
are shown in Figs.~\ref{f5}a,b [using the parameter sets (a) and
(b) respectively]. We also show separately the contributions to the 
amplitude from the quark and the hadron molecule components (dashed and 
dashed-dotted curves, respectively). The comparison with the standard 
$^3P_0$ model calculation with a fixed value for the vector-meson radius 
$b_V=0.9\,b$ (dotted curves) demonstrates the following: a smooth transition 
from the $^3P_0$ $\gamma RN$ vertex (Fig.~\ref{f1}a) to the parton-like one 
(Fig.~\ref{f1}b) using a $Q^2$-dependent vector meson radius 
$b_V(Q^2)\to\,$0 leads to considerable improvement of the standard $^3P_0$ 
model results at moderate values of $Q^2$. 

The quark core component of R plays the main role in the  electroproduction of 
the Roper resonance for this $Q^2$ region ($Q^2\gtrsim\,$1 -- 1.5 GeV$^2/c^2$). 
For small values of $Q^2\lesssim\,$1 GeV$^2$, where the contribution of the 
meson cloud should also be important, it can be effectively taken into account 
in the framework of $^3P_0$- and VMD models. However, such a model 
overestimates the transverse amplitude $A_{1/2}$ in the region 
0.5$\,\lesssim Q^2\lesssim\,$1 GeV$^2$ (the dashed line in Fig.~\ref{f5}). The 
description of the JLab data~\cite{aznauryan09,mokeev09} on $A_{1/2}$ can be 
considerably improved if one takes a combined structure for the Roper in the 
form of $|R\ra=\cos\theta|3q^\ast\rangle+\sin\theta|N+\sigma\ra$. 
The adjustable parameter $\theta$ fitted to the JLab data in this region is 
about  $\cos\theta=\,$0.8 [for the `hard` variant (a)] or $\cos\theta=\,$0.7 
[for set (b)], in both cases indicating an admixture of $N\sigma$ component of
about 50\%.

The `hard` version (a) looks more plausible in the description of both 
amplitudes $A_{1/2}$ and $S_{1/2}$, while set (b) only represents a fit to the 
soft-$Q^2$ (up to $Q^2\approx\,$1 GeV$^2/c^2$) behavior of the transverse 
amplitude $A_{1/2}$. In the soft-$Q^2$ region the contribution of the pion 
cloud and the influence of the coupled channel $\Delta+\pi$ are 
importantd~\cite{mokeev09,golli09,matsuyama07,suzuki09,kamano10}. Both 
effects should be taken into account in further detailed calculations. 

\subsection{Decay widths}
\begin{figure}[hp]
\begin{center}
\epsfig{figure=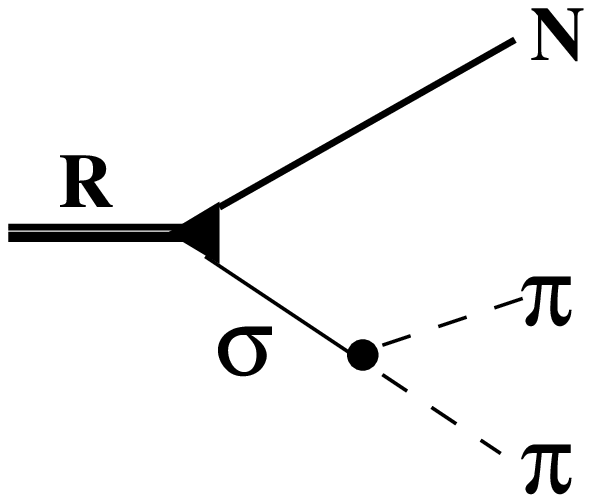,width=0.17\textwidth,clip}
\end{center}
\caption{$R\to N+(\pi+\pi)_{I=0}$} decay process via the 
$\sigma$-meson resonance.
\label{f7}
\end{figure}

When the weight of $N+\sigma$ component in the Roper resonance in terms of
$\sim\sin^2{\theta}$ is fixed, the Roper decay width for the transition
$N+(\pi\pi)_{Swave}^{I=0}$ can be calculated.
The assumption that the quark part of the Roper just gives a very small 
contribution through a virtual transition $R\to N+\sigma$ is justified in our
quark model [see, e.g., our evaluation of the quark amplitude 
${\cal M}^q_{R\to N+\sigma}$ in Eq.~(\ref{a31}) which goes to zero at 
$y_\sigma=1$ as it follows from Eq.~(\ref{a30z})]. Then the transition is 
described as the virtual decay of the molecular part to $N+\sigma$ 
followed by the $\sigma\to\pi\pi$ decay. The diagram for such a mechanism is 
shown in Fig.~\ref{f7}.

The probability $\vert M_{fi}\vert^2$ for the transition process of 
Fig.~\ref{f7} contains the Breit-Wigner representation for the intermediate 
$\sigma$-meson state with
\eq
\vert M_{fi}\vert^2&=&
g_{{\scriptscriptstyle R}\sigma {\scriptscriptstyle N}}^2\,
g_{\sigma\pi\pi}^2 \tilde\Phi_R^2(k^2)
\frac{(m_N+m_R)^2-s_{\pi\pi}}{(m_\sigma^2-s_{\pi\pi})^2+m_\sigma^2\,
\Gamma_\sigma^2(s_{\pi\pi})},\nonumber\\
\Gamma_\sigma(x)&=&\Gamma_\sigma\frac{m_\sigma}{\sqrt{x}}
\frac{\sqrt{x-4m_\pi^2}}{\sqrt{m_\sigma^2-4m_\pi^2}},\quad 
x=s_{\pi\pi}\equiv k_\sigma^2\,, 
\label{c1}
\en
where 
$k=p_R-\omega_{N\sigma}\,p_N$ and the coupling constant $g_{\sigma\pi\pi}$ 
is deduced from the two-pion decay width of the $\sigma$ with 
$g_{\sigma\pi\pi}^2=\frac{32\pi}{3}
\Gamma_\sigma\,m_\sigma \, \Big(1-\frac{4m_\pi^2}{m_\sigma^2}\Big)^{-1/2}$. 
The coupling constant $g_{{\scriptscriptstyle R}\sigma {\scriptscriptstyle N}}$
of the hadron-molecular vertex is defined by the compositeness 
condition~(\ref{a33}).

The result for the $R\to N+(\pi\pi)_{Swave}^{I=0}$ decay width is presented 
by an integral over the variables of the phase space volume
\eq
\label{width_RhoNPiPi}
\Gamma_{R\to N\sigma(\pi\pi)}&=&
\frac{3\, \sin^2\theta}{512\pi^3\,m_R^3}
\int\limits_{4m_\pi^2}^{(m_R-m_N)^2} 
\frac{ds_{\pi\pi}}{s_{\pi\pi}}\, \vert M_{fi}\vert^2 \label{c2}\\
&\times&\lambda^{1/2}(m_R^2,m_N^2,s_{\pi\pi})\,
\lambda^{1/2}(s_{\pi\pi},m_\pi^2,m_\pi^2) 
\nonumber
\en
with $\lambda(a,b,c)=a^2+b^2+c^2-2ab-2bc-2ac$.

The numerical value for $\Gamma_{R\to N\sigma}$ with $g_{R\sigma N}=6.39$
[fixed by the compositeness condition (\ref{a33})] and a molecular admixture 
in the Roper of $\sin{\theta}\simeq 0.6$ is 
\eq
\Gamma_{R\to N\sigma(\pi\pi)}= ( 19.0 \,- \, 26.7) \ {\rm MeV}\,.
\label{c3}
\en 
where the lower and upper limits correspond to a variation of
the $\sigma $ decay width $\Gamma_\sigma$ from 0.5 to 1 GeV, respectively 
(the variation of the $\sigma$-meson mass $M_\sigma=500\pm 50\,$MeV can only 
change the result within 10\%).
This should be compared to the PDG~\cite{pdg10} value \\
$\Gamma_{R\to N\sigma(\pi\pi)}\approx\,$(0.05 -- 0.1)$\,\Gamma_R^{\rm tot}$
($\approx 15 - 30\,$MeV) or the
recent data~\cite{sarantsev08} $\Gamma_{R\to N\sigma}=\,$ 71$\,\pm\,$17 MeV. 
It is clear that the strong Roper 
decay can serve as a constraint on $\Gamma_\sigma$, however present 
results for $\Gamma_{R\to N\sigma(\pi\pi)}$ are compatible with all 
values of $\Gamma_\sigma$. 
 
The pion decay width calculated for the quark part of the Roper resonance
in the framework of our approach ($\Gamma^q_{R\to\pi N}\simeq\,$36 MeV) is not 
as small as in the case of EEM evaluations 
($\Gamma^{\scriptscriptstyle EEM}_{R\to\pi N}\simeq\,$4 MeV) but it is still 
several times smaller than the PDG value of 
$\Gamma_{R\to N+\pi}\approx\,$(0.55 -- 0.75)$\,\Gamma_R^{\rm tot}$.
It is clear that considerable corrections to $\Gamma^q_{R\to\pi N}$ can come 
from the pion cloud contribution which is neglected here. 

\section{Conclusions}

We suggested a two-component model of the lightest nucleon resonance
$R=N_{1/2^+}$(1440) as a combined state of the quark configuration $sp^2[3]_X$
and the hadron molecule component $N+\sigma$. This approach allows to describe 
with reasonable accuracy the recent CLAS electroproduction 
data~\cite{aznauryan09,mokeev09} at low- and moderate values of $Q^2$ with
0$\,\le Q^2\lesssim\,$1.5 -- 2 GeV$^2$. In the model the 
$R\to N+(\pi\pi)^{I=0}_{Swave}$ transition process is interpreted as the decay 
of a virtual $\sigma$ meson in the $N+\sigma$ component. The calculated decay 
width $\Gamma_{R\to N\sigma(\pi\pi)}$ correlates well with the PDG 
value~\cite{pdg10} and the recent CB-ELSA and A2-TAPS data~\cite{sarantsev08}.

The weight of the $N+\sigma$ component in the Roper with 
$\sin^2\theta\approx$ 0.36 is compatible with the CLAS data at low and 
moderate $Q^2$. This weight is also compatible with the value of the
helicity amplitude $A_{1/2}$ at the photon point and with the  
data on the $R\to N+(\pi\pi)^{I=0}_{Swave}$ decay width. 

However, our evaluations have 
shown that at low $Q^2$ the contribution of the pion cloud to the amplitude 
$A_{1/2}$ can be considerable. For example, this is evident from 
Fig.~\ref{f5}a, where the discrepancy  of our results and the CLAS data is 
about 1 -- 1.5 experimental error bars. Still, this discrepancy is considerably 
smaller than in the case of previous quark models: note the 
predictions of the valence quark covariant spectator model (the 
dashed-double-dotted curve in Fig.~\ref{f5}a adapted from 
Ref.~\cite{romalho10}) or predictions of the LF models in the same region 
of $Q^2\lesssim\,$ 1 GeV$^2/c^2$.

In this paper we tried to show that the description of transition amplitudes 
in terms of parton-like models, which are very good at high $Q^2$, can be 
naturally transformed into a description in terms of the `soft` vector meson 
cloud. This smooth transition is achieved by 'switching on' a non-zero radius 
of the intermediate vector meson. The vector meson $V$ of finite size 
generates a non-local $Vqq$ interaction. 
This weakens the effect of the orthogonality of the spatial $R$ and $N$ wave 
functions in the transition matrix element $N+\gamma_T^*\to R$,
and the amplitude $A_{1/2}$. Resulting theoretical values, which match the 
data, are contrary to the standard predictions of LF-models, which lead to
non-zero and (negative) large values at the photon point. 

Further we plan to develop a relativistic version of the suggested 
electroexcitation mechanism. 

\begin{acknowledgments}  
The authors would like to thank Prof. V.I. Mokeev for presentation of the new 
essential information on analysis of recent JLab data. We would also like 
to thank Profs. V.I. Kukulin and V.G. Neudatchin for stimulating discussions.
This work was supported by the DFG under Contract
No. FA67/31-2. 
The work is partially supported by the DFG under Contract No. 
436 RUS 113/988/01 and by the grant No. 09-02-91344 of RFBR (the Russian 
Foundation for Basic Research). This research is also part of the 
European Community-Research Infrastructure Integrating Activity 
``Study of Strongly Interacting Matter'' (HadronPhysics2, 
Grant Agreement No. 227431) and Federal Targeted Program "Scientific and 
scientific-pedagogical personnel of innovative Russia" 
Contract No. 02.740.11.0238.
\end{acknowledgments}


\appendix

\section{Hadron quark wave functions}
\subsection{Basic elements}
We use the standard definitions for the harmonic oscillator wave functions
\eq
\varphi_{0S}({\bf p},r_0)&=&(4\pi r_0^2)^{3/4}e^{-p^2r_0^2/2}\,,\label{a1} \\
\varphi_{2S}({\bf p},r_0)&=&\sqrt{\frac{3}{2}}(1-\frac{2}{3}p^2r_0^2)
\varphi_{0S}({\bf p},r_0)\,,\nonumber\\
\varphi_{1P,m}({\bf p},r_0)&=&\sqrt{\frac{2}{3}}p\,r_0\varphi_{0S}({\bf p},r_0)
\sqrt{4\pi}Y_{1m}(\hat{\bf p})\,.\nonumber
\en 
The relative momenta in the quark (antiquark) systems with numbering 
i=1,2,3,4,5 (see Figs.~\ref{f1} and \ref{f6}) are set up as
\begin{eqnarray}
\bm{\varkappa}_1&=&\frac{1}{2}({\bf p}_1-{\bf p}_2), \ 
\bm{\varkappa}_2=\frac{1}{3}({\bf p}_1+{\bf p}_2)
-\frac{2}{3}{\bf p}_4,\nonumber\\
\bm{\varkappa}^\prime_2&=&\frac{1}{3}({\bf p}_1+{\bf p}_2)
-\frac{2}{3}{\bf p}_3, \label{a2}\\
\b{\varkappa}_{\scriptscriptstyle M}&=&
\frac{1}{2}({\bf p}_3-{\bf p}_5), \
{\bf p}_{\scriptscriptstyle M}={\bf p}_3+{\bf p}_5
\nonumber
\en 
In the rest frame with
${\bf P}^\prime_{\scriptscriptstyle R(N)}={\bf p}_1+{\bf p}_2+{\bf p}_3=0$ 
we have the relations
\eq
{\bf p}_{\scriptscriptstyle M}&=&{\bf k}, \
{\bf P}_{\scriptscriptstyle N}={\bf p}_1+{\bf p}_2+{\bf p}_4=-{\bf k}, \
{\bf p}_4-{\bf p}_3=-{\bf k},\nonumber \\
\b{\varkappa}_{\scriptscriptstyle M}&=&-\bm{\varkappa}^\prime_2-\frac{\bf k}{2}, \
{\bf p}_4+{\bf p}_3=-2\bm{\varkappa}^\prime_2-{\bf k}
\label{a2b}
\en
which are used with $m_1=m_2=m_3=m_q=m_N/3$ and
\eq
& &\nu_1=2, \ \nu_2=\frac{3}{2}, \
m_{12}=\frac{m_1m_2}{m_1+m_2}=\frac{m_q}{\nu_1}, \nonumber\\  
& &m_{(12)3}=\frac{(m_1+m_2)\,m_3}{m_1+m_2+m_3}=\frac{m_q}{\nu_2}, 
\label{a2c}
\en
in the calculation of matrix elements.
\subsection{Quark configurations}\label{ap1}
\subsubsection{Baryons}
The translationally invariant quark configurations \\ $s^3[3]_X$ and
$sp^2[3]_x,L=0$ are represented in terms of harmonic oscillator wave functions
(\ref{a1}) depending on the relative momenta (\ref{a2}) as
\eq 
& &
\Psi_N(\b{\varkappa}_1,\b{\varkappa}_2;{\scriptstyle S}_z,{\scriptstyle T}_z)
\nonumber\\
&=&\varphi_{0S}(\b{\varkappa}_1,\sqrt{\nu_1}b)\,
\varphi_{0S}(\b{\varkappa}_2,\sqrt{\nu_2}b)\psi_N^{ST}(124)\,, 
\en 
\eq 
& &\Psi_R(\b{\varkappa}_1,\b{\varkappa}^\prime_2;
{\scriptstyle S}_z,{\scriptstyle T}_z)\nonumber\\
&=&\biggl[\sqrt{\frac{1}{2}}
\varphi_{0S}(\b{\varkappa}_1,\sqrt{\nu_1}b)\,
\varphi_{2S}(\b{\varkappa}^\prime_2,\sqrt{\nu_2}b) \\
&+&\sqrt{\frac{1}{2}}
\varphi_{2S}(\b{\varkappa}_1,\sqrt{\nu_1}b)\,
\varphi_{0S}(\b{\varkappa}^\prime_2,\sqrt{\nu_2}b)
\biggr]\psi_N^{ST}(123). \nonumber
\label{a4}
\en
The spin-isospin part $\psi_N^{ST}$ for both configurations is 
described by the state vector
\eq 
& &\psi_N^{ST}(124)
=\sum_{\mu_4t_4}\!\biggl[\sqrt{\frac{1}{2}}
(1({\scriptstyle S}_z\!-\!\mu_4){\scriptstyle\frac{1}{2}}\mu_4|
{\scriptstyle\frac{1}{2}}{\scriptstyle S}_z)\nonumber\\
&\times&\!(1({\scriptstyle T}_z\!-\!t_4){\scriptstyle\frac{1}{2}}t_4|
{\scriptstyle\frac{1}{2}}{\scriptstyle T}_z)\,|{\scriptstyle S}_{12}\!=\!1,\,
{\scriptstyle S}_z\!-\!\mu_4\rangle\,|{\scriptstyle T}_{12}\!=\!1,\,
{\scriptstyle T}_z\!-\!t_4\rangle\nonumber\\
&+&\!\!\sqrt{\frac{1}{2}}\delta_{\mu_4,\,{\scriptstyle S}_z}
\delta_{t_4,\,{\scriptstyle T}_z}|{\scriptstyle S}_{12}\!=\!0,0\rangle\,
|{\scriptstyle T}_{12}\!=\!0,0\rangle\biggr]\,\chi_{\mu_4}\xi_{t_4},
\label{a3}
\en
where $\chi_{\mu_i}$ ($\xi_{t_i}$) is the spinor (isospinor) of i-th quark,
 $\mu_i$ ($t_i$) is the spin (isospin) projection; i=1,2,4 for the nucleon
and i=1,2,3 for the Roper.
\subsubsection{Mesons}
1) Pseudoscalar ($\pi$) and scalar ($\sigma$)
\eq 
\Psi_\pi(\b{\varkappa}_{\pi}, t_{\pi})&=&
\varphi_{0S}(\b{\varkappa}_{\pi},\sqrt{\nu_1}b_{\pi})\sum_{\mu_3\mu5}
({\scriptstyle\frac{1}{2}}\mu_3\,{\scriptstyle\frac{1}{2}}\mu_5|00)
\nonumber\\
&\times&\!
\sum_{t_3t_5}({\scriptstyle\frac{1}{2}}t_3\,{\scriptstyle\frac{1}{2}}t_5
|1t_\pi)\,\chi_{\mu_3}\chi_{\mu_5}\xi_{t_3}\xi_{t_5}\qquad\quad
\label{a7}
\en
\eq 
& &\Psi_\sigma(\b{\varkappa}_\sigma)
=\varphi_{0S}(\b{\varkappa}_{\sigma},\sqrt{\nu_1}b_{\sigma})\,
\sqrt{\nu_1}b_{\sigma}\varkappa_{\sigma}\nonumber\\
&\times&\!\sqrt{\frac{4\pi}{3}}
\sum_m(1m1-\!m|00)\sum_{\mu_3\mu_5}
({\scriptstyle\frac{1}{2}}\mu_3\,{\scriptstyle\frac{1}{2}}\mu_5|1m)
Y_{1-m}(\hat{\b{\varkappa}_\sigma})\nonumber\\
&\times&
\sum_{t_3t_5}({\scriptstyle\frac{1}{2}}t_3\,{\scriptstyle\frac{1}{2}}t_5|00)
)\,\chi_{\mu_3}\chi_{\mu_5}\xi_{t_3}\xi_{t_5}
\label{a6}
\en

2) Vectors ($\rho,\,\omega$)
\eq 
\Psi_\rho(\b{\varkappa}_{\rho},\lambda_{\rho},t_{\rho})&=&
\varphi_{0S}(\b{\varkappa}_{\rho},\sqrt{\nu_1}b_{\rho})\sum_{\mu_3\mu5}
({\scriptstyle\frac{1}{2}}\mu_3\,{\scriptstyle\frac{1}{2}}\mu_5|1\lambda_\rho)
\nonumber\\
&\times&\! 
\sum_{t_3t_5}({\scriptstyle\frac{1}{2}}t_3\,{\scriptstyle\frac{1}{2}}t_5
|1t_\rho)\,\chi_{\mu_3}\chi_{\mu_5}\xi_{t_3}\xi_{t_5}
\label{a5}
\en 
(for the $\omega$ use the substitution 
\eq 
({\scriptstyle\frac{1}{2}}t_3\,{\scriptstyle\frac{1}{2}}t_5|1t_\rho)\to
({\scriptstyle\frac{1}{2}}t_3\,{\scriptstyle\frac{1}{2}}t_5|00) 
\en 
in Eq.~(\ref{a5})).

 \section{Meson-baryon coupling}\label{ap2}
The meson-baryon vertex generated by the effective 
pair-creation operator 
$V_{\bar qq}^{eff}$ is schematically sketched in Fig.~\ref{f6}.
Performing a re\-co\-up\-ling of  
qu\-ark (anti\-quark) variables in the matrix
element  $\langle M|\langle N|V_{\bar qq}^{eff}|N(R)\rangle$ 
($M=\pi,\,\sigma,\,\rho,\,\omega$), omitting the isospin part and other 
trivial factors ($\frac{g_q}{2m_q}$, 
$\frac{g_q}{2m_q}$, $\delta({\bf p}_4+{\bf p}_5)$, etc.) we obtain for the
non-trivial spin part of the effective $\pi qq$ and $\rho qq$ vertices the 
following expressions:
\eq
\pi qq:\,\,{\bm \sigma\cdot}{\bf p}_4,\quad \rho qq:\,\,
(-{\bf p}_4+i[{\bm\sigma\times}{\bf p}_4])\!\cdot\!\b{\epsilon}^{(\lambda_\rho)},
\label{apr}
\en
where 
\eq
\epsilon^{(\lambda_\rho)\mu}=\bigl\{\epsilon_0^{(\lambda_\rho)},
\b{\epsilon}^{(\lambda_\rho)}\bigr\}
\label{aer}
\en
is the $\rho$ meson polarization vector and ${\bm \sigma}$ is the vector of 
quark spin Pauli matrices. Expressions in Eq.~(\ref{apr}) are only acceptable 
in the rest frame of the initial baryon $N(R)$, in which case 
${\bf p}_4=-{\bf k}$ and ${\bf p}_4=-{\bf p}_5$ (see Fig~\ref{f6}). For the 
3rd quark with a non-zero momentum ${\bf p}_3\not=\,$0 both expressions do not 
satisfy the Galilean invariance and the second expression in Eq.~(\ref{apr}) 
does not correspond to the elementary $\rho qq$ vertex 
$\bar u(p_4)\gamma^\mu u(p_3)\rho_\mu(p_3-p_4)$. 
Thus it does not lead to a conserved current in the VMD model.
\begin{figure}[hp]
\begin{center}
\epsfig{figure=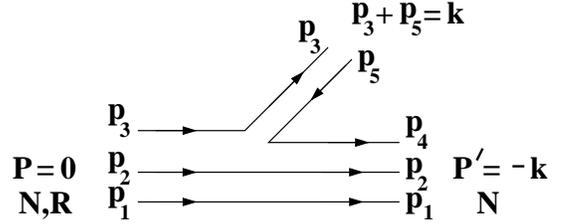,width=0.4\textwidth,clip}
\end{center}
\caption{Quark diagram of the $^3P_0$ model for the meson-baryon coupling.}
\label{f6}
\end{figure}

It is possible to improve the expressions (\ref{apr}) in an acceptable
form without changing them in the rest frame where they were deduced in the 
$^3P_0$ ansatz. Such corrections are only possible with the substitutions 
${\bf p}_4\to{\bf p}_4\pm{\bf p}_3$, which become identities for 
${\bf p}_3=\,$0, i.e. in the rest frame of the 3-rd spectator quark. 

In our calculations we use the following corrected form of Eq.~(\ref{apr}): 
\eq
\pi qq:& & {\bm \sigma\cdot}({\bf p}_4-{\bf p}_3),\label{a7a}\\
\rho qq:& & (E_4+E_3)\epsilon_0^{(\lambda_\rho)}\nonumber\\
&-&\bigl\{({\bf p}_4+{\bf p}_3)-i[{\bm\sigma\times}({\bf p}_4-{\bf p}_3)]\bigr\}
\!\cdot\!\b{\epsilon}^{(\lambda_\rho)}\bigr).\nonumber
\en
These expressions satisfy Galilean invariance \\
and are well correlated with the Feynman amplitudes \\ 
$\bar u(p_4)\gamma^5 u(p_3)\pi(p_3-p_4)$ and
$\bar u(p_4)\gamma^\mu u(p_3)\rho_\mu(p_3-p_4)$, respectively (in the 
non-relativistic approximation). Here we show that using
such corrected form of Eq.~(\ref{apr}) one can obtain realistic values for the
coupling constants $\rho NN$, $\omega NN$, $\sigma NN$ and the nucleon 
magnetic moments $\mu_p,\,\mu_n$. We further predict the non-diagonal 
couplings $\pi NR$, $\sigma NR$, $\rho NR$, $\omega NR$ starting from a single 
constant $\gamma=\frac{g_q}{2m_q}$ normalized to the well established value
$g_{\pi{\scriptscriptstyle NN}}=\,$13.5.

\subsection{Diagonal $N\to N$ transitions}
Substituting wave functions (\ref{a7}) -- (\ref{a5}) into Eq.~(\ref{a36a})
and taking into account the modification (\ref{a7a}) of Eq.~(\ref{apr})
gives after some algebra the following expressions for the 
$N\to N+M$ ($N+M\to N$) amplitudes:
\eq
& &T^q_{N\to N+\pi}=3\,_{nr}\!\langle\Psi_\pi,{\bf k}|\langle\Psi_N,-{\bf k}|
V^{eff}_q|\Psi_N,{\bf 0}\rangle_{nr}\nonumber\\
&=&\frac{5}{3}\!\left(\!\frac{g_q}{2m_q}\!\right)\frac{1}{n^{3/2}(y_\pi)}\,
(8\pi b^2)^{3/4}y^{3/2}_\pi e^{-\zeta(y_\pi){\bf k}^2b^2/6}\nonumber\\
&\times&
\langle{\scriptstyle\frac{1}{2}}{\scriptstyle T}^\prime_z|\tau_{t_\pi}|
{\scriptstyle\frac{1}{2}}{\scriptstyle T}_z\rangle
\langle{\scriptstyle\frac{1}{2}}{\scriptstyle S}^\prime_z|
\frac{{\bm\sigma}\!\cdot\!{\bf k}}{2m_q}|
{\scriptstyle\frac{1}{2}}{\scriptstyle S}_z\rangle,
\label{a13}
\en
\eq
& &T^q_{N\to N+\sigma}=3\,_{nr}\!\langle\Psi_\sigma,{\bf k}|\langle\Psi_N,-{\bf k}|
V^{eff}_q|\Psi_N,{\bf 0}\rangle_{nr}\nonumber\\
&=&\!-3\left(\!\frac{g_q}{2m_q}\!\right)\!\frac{y_\sigma}{\sqrt{6}(2m_qb)}\,
\frac{1}{n^{5/2}(y_\sigma)}(8\pi b^2)^{3/4}y^{3/2}_\sigma\,\nonumber\\
&\times&\!\!\left(1+f(y_\sigma)\,\frac{{\bf k}^2b^2}{9}
\!\right)
e^{-\zeta(y_\sigma){\bf k}^2b^2/6}\delta_{{\scriptstyle S}^\prime_z,{\scriptstyle S}_z}
\delta_{{\scriptstyle T}^\prime_z,{\scriptstyle T}_z},
\label{a14}
\en
\eq
&&T^q_{\rho+N\to N}=3\,_{nr}\!\langle\Psi_N,{\bf 0}|V^{eff}_q
|\Psi_N,-{\bf k}\rangle|\Psi_\rho,{\bf k},\lambda_\rho,t_\rho\rangle_{nr}
\nonumber\\
&&=\!-\left(\!\frac{g_q}{2m_q}\!\right)\,
\frac{1}{n^{3/2}(y_\rho)}(8\pi b^2)^{3/4}y^{3/2}_\rho 
e^{-\zeta(y_\rho){\bf k}^2b^2/6}\nonumber\\
&&\times\langle{\scriptstyle\frac{1}{2}}{\scriptstyle T}^\prime_z
|\tau_{t_\rho}|{\scriptstyle\frac{1}{2}}{\scriptstyle T}_z\rangle
\biggl[\!\bigl(1+
\frac{1}{3}\frac{{\bf k}\!\cdot\!{\bm\epsilon}^{(\lambda_\rho)}}{2m_qn(y_\rho)}
\bigr)\delta_{{\scriptstyle S}^\prime_z,{\scriptstyle S}_z}\nonumber\\
&&-\frac{5}{3}
\langle{\scriptstyle\frac{1}{2}}{\scriptstyle S}^\prime_z|
\frac{i[{\bm\sigma}\times{\bf k}]\!
\cdot\!{\bm\epsilon}^{(\lambda_\rho)}}{2m_q}|
{\scriptstyle\frac{1}{2}}{\scriptstyle S}_z\rangle
\biggr],
\label{a15}
\en
\eq
T^q_{\omega+N\to N}=3\,_{nr}\!\langle\Psi_N,{\bf 0}|V^{eff}_q
|\Psi_N,-{\bf k}\rangle|\Psi_\omega,{\bf k},\lambda_\omega\rangle_{nr}\quad
\nonumber\\
=-\!\left(\!\frac{g_q}{2m_q}\!\right)\!\frac{1)}{n^{3/2}(y_\omega)}
(8\pi b^2)^{3/4}y^{3/2}_\omega e^{-\zeta(y_\omega)k^2b^2/6}\quad\\
\times\delta_{{\scriptstyle T}^\prime_z,{\scriptstyle T}_z}\biggl[\!\bigl(
3+\frac{{\bf k}\!\cdot\!{\bm\epsilon}^{(\lambda_\omega)}}{2m_qn(y_\omega)}\bigr)
\delta_{{\scriptstyle S}^\prime_z,{\scriptstyle S}_z}\!
-\langle
\frac{i[{\bm\sigma}\times{\bf k}]\!
\cdot\!{\bm\epsilon}^{(\lambda_\omega)}}{2m_q}\rangle_S\!
\biggr].\nonumber
\label{a16}
\en
The parameter $b$ is the r.m.s. radius of the quark configuration $0s^3$ which
is used for the nucleon. The meson radius $b_{\scriptstyle M}$ is related through
the relative value
\eq
y_{\scriptscriptstyle M}=\frac{b_{\scriptscriptstyle M}}{b},\qquad
M=\pi,\,\sigma,\,\rho,\,...,
\label{a17}
\en
and we use the notations
\eq
n(y)&=&1+\frac{2}{3}y^2,\quad\zeta(y)=\frac{1+5y^2/6}{n(y)},\nonumber\\
f(y)&=&\frac{1\!+\!y^2/2}{n(y)}.
\label{a17a}
\en
If $b_{\scriptscriptstyle R}\not=b$ we also use another relative variable
$y_{\scriptscriptstyle R}=\frac{b_{\scriptscriptstyle R}}{b}$ and then 
Eq.~(\ref{a17}) should be generalized as
\eq
n(y)\to N(y,y_{\scriptscriptstyle R})=\frac{1}{2}(1+y^2_{\scriptscriptstyle R})
+\frac{2}{3}y.
\label{a17b}
\en
The strength parameter of the $^3P_0$ model, $\gamma=\frac{g_q}{2m_q}$, is 
fixed as usual  by normalizing the value of the $\pi NN$ coupling constant to
$g_{\pi{\scriptstyle NN}}=13.5$. From (\ref{a13}) it follows that
\eq
g^q_{\pi{\scriptscriptstyle NN}}&=&\frac{5}{3}\frac{m_N}{m_q}
\left(\frac{g_q}{2m_q}\right)\frac{1}{n^{3/2}(y_\pi)}\,
\nonumber\\
&\times&\left[(8\pi b^2)^{3/4}y^{3/2}_\pi
2m_N\sqrt{2M_\pi}\right]
\label{a24}
\en
and for a typical value of $b=$ 0.5 fm one obtains $\gamma\simeq\,$ 0.2.

For the $\sigma NN$ coupling constant
\eq
g^q_{\sigma{\scriptscriptstyle NN}}&=&
3\left(\!\frac{g_q}{2m_q}\!\right)\frac{y_\sigma}{\sqrt{6}(2m_qb)}
\frac{1}{n^{5/2}(y_\sigma)}\nonumber\\
&\times&\left[(8\pi b^2)^{3/4}y^{3/2}_\sigma2m_N\sqrt{2M_\sigma}\right]
\label{a27}
\en
we get
\eq
g^q_{\sigma{\scriptscriptstyle NN}}=
\frac{g_{\pi{\scriptstyle NN}}}{2m_Nb}\frac{9\sqrt{3}}{10\sqrt{2}}
\sqrt{\frac{M_\sigma}{M_\pi}}
\frac{y^{5/2}_\sigma}{y^{3/2}_\pi}\,\frac{n^{3/2}(y_\pi)}{n^{5/2}(y_\sigma)}.\quad
\label{a27a}
\en
taking $g^q_{\pi{\scriptscriptstyle NN}}=g_{\pi{\scriptscriptstyle NN}}$.
For typical CQM values of $b=$ 0.5$\,$fm and  $y_\sigma=y_\pi=$ 1 this 
expression gives a realistic value for the coupling constant with
$g^q_{\sigma{\scriptscriptstyle NN}}=0.262g_{\pi{\scriptscriptstyle NN}}=$ 3.54.

For the $\rho NN$ coupling constant defined in Eq.~(\ref{a15}) as
\begin{equation}
g^q_{\rho{\scriptscriptstyle NN}}=\frac{1}{3}
\left(\!\frac{g_q}{2m_q}\!\right)
\frac{1}{n^{3/2}(y)}\left[(8\pi b^2)^{3/4}y^{3/2}2m_N\sqrt{2M_V}
\right]
\label{a23}
\end{equation}
substitution of the value $\frac{g_q}{2m_q}$ deduced from Eq.~(\ref{a24})
gives
\begin{equation}
g^q_{\rho{\scriptscriptstyle NN}}=\frac{g_{\pi{\scriptscriptstyle NN}}}{5}
\sqrt{\frac{M_\rho}{M_\pi}}\frac{y^{3/2}_\rho}{y^{3/2}_\pi}\,
\frac{n^{3/2}(y_\pi)}{n^{3/2}(y_\rho)}
\label{a25}
\end{equation}
and for $y_\rho=y_\pi=$ 1 one obtains the realistic value 
$g_{\rho{\scriptstyle NN}}=g^q_{\rho{\scriptstyle NN}}=
0.469g_{\pi{\scriptstyle NN}}=$ 6.33.

Comparing Eqs.~(\ref{a15}) and (\ref{a16}) one can see that in this approach 
the $\omega NN$- and $\rho NN$ couplings are linked by the standard relation
\begin{equation}
g^q_{\omega{\scriptscriptstyle NN}}=3g^q_{\rho{\scriptscriptstyle NN}} 
\label{a25a}
\end{equation}
which corresponds to ``ideal mixing'' usually used in the VMD model. 
\subsection{Non-diagonal transitions}
Here the main objective is the calculation of the non-diagonal baryon matrix 
elements for the transitions $N+\rho\to R$ and $R\to N+M$.
The values of the coupling constants have been fixed by Eqs.~(\ref{a27a}),
(\ref{a25}) and (\ref{a25a}) on the basis of $g_{\pi{\scriptscriptstyle NN}}$. 
We further use them in the expressions for the
non-diagonal transitions $N+\gamma^*\to R$, $R\to N+\pi$, $R\to N+\sigma$, etc.
substituting symbols $g^q_{\sigma{\scriptscriptstyle NN}}$ and 
$g^q_{\rho{\scriptscriptstyle NN}}$ (and $g^q_{\omega{\scriptscriptstyle NN}}$ with 
$g^q_{\omega{\scriptscriptstyle NN}}=3g^q_{\rho{\scriptscriptstyle NN}}$) instead of 
the explicit expressions of the r.h.s. of Eqs.~(\ref{a27}) and (\ref{a23}). 
Then the vector meson absorption amplitude 
${\cal M}^{q(\lambda_V)}_{V+N\to R}=2m_N\sqrt{2M_V}T^{q(\lambda_V)}_{V+N\to R}$ 
is represented by the following two-component column vector: 
\eq
\left(\!\mbox{\begin{tabular}{c} ${\cal M}^{q(\lambda_\rho)}_{\rho+N\to R}$\\
${\cal M}^{q(\lambda_\omega)}_{\omega+N\to R}$\end{tabular}}\!\right)\!=
\frac{\sqrt{3}}{2}g^q_{\rho{\scriptstyle NN}}e^{-\zeta(y)k^2b^2/6}\!
\left(\!\!\mbox{\begin{tabular}{c} 
$\langle{\scriptstyle\frac{1}{2}}{\scriptstyle T}^\prime_z|\tau_{t_\rho}|
{\scriptstyle\frac{1}{2}}{\scriptstyle T}_z\rangle$\\
3$\delta_{{\scriptscriptstyle T}^\prime_z,{\scriptscriptstyle T}_z}$\end{tabular}}\!\!
\right)
\nonumber\\
\times\!\biggl\{\!\!\biggl[\!\bigl(\!\epsilon_0^{(\lambda)}\!+
\frac{n_0{\bf \tilde k}\!\cdot\!\bm{\epsilon}^{(\lambda)}}{2m_Nn(y)}\bigr)
p_2(y,{\bf k}^2)\!
+p_0(y)\frac{n_0{\bf \tilde k}\!\cdot\!\bm{\epsilon}^{(\lambda)}}{2m_Nn(y)}\biggr]
\delta_{{\scriptscriptstyle S}^\prime_z,{\scriptscriptstyle S}_z}\nonumber\\
+\left(\!\mbox{\begin{tabular}{c}$5$\\$1$\end{tabular}}\!\right)\!
\langle{\scriptstyle\frac{1}{2}}{\scriptstyle S}^\prime_z|
\frac{i[{\bm\sigma}\times{\bf k}]\!\cdot\!\epsilon^{(\lambda)}}{2m_N}
|{\scriptstyle\frac{1}{2}}{\scriptstyle S}_z
\rangle p_2(y,{\bf k}^2)\!
\biggr\}.\qquad\qquad
\label{a28}
\en 
This is the main result of our considerations. Here we use momenta 
${\bf k}={\bf P}-{\bf P}^\prime$, ${\bf \tilde k}={\bf P}+{\bf P}^\prime$, 
related to momenta ${\bf P}$, ${\bf P}^\prime$ of initial and final baryon.
Only in the rest frame they have
the same values, ${\bf \tilde k}={\bf k}$. In the case 
$b_{\scriptscriptstyle R}\not=b$ the polynomials $p_{0,2}$ and $\zeta,n,n_0$ also
depend on $y_{\scriptscriptstyle R}$. They are defined by the equations
\eq
p_0(y)=P_0(y,y_{\scriptscriptstyle R}=1),\,\,
p_2(y,{\bf q}^2)=P_2(y,y_{\scriptscriptstyle R}=1,{\bf q}^2),\nonumber\\
n_0=N_0(y_{\scriptscriptstyle R}=1),\,\,N_0(y_{\scriptscriptstyle R})=
\frac{3}{2}y^2_{\scriptscriptstyle R}-\frac{1}{2},\qquad\qquad\nonumber\qquad\\
\zeta(y)=\tilde\zeta(y,y_{\scriptscriptstyle R}=1),\,\,
\tilde\zeta(y,y_{\scriptscriptstyle R})=
\frac{y^2_{\scriptscriptstyle R}+\frac{3}{2}(\frac{1+y^2_{\scriptscriptstyle R}}{2}
-\frac{4}{9})y^2}{\frac{1+y^2_{\scriptscriptstyle R}}{2}+\frac{2}{3}y^2},
\nonumber\\
P_0(y,y_{\scriptscriptstyle R})=
\frac{4}{3}\frac{1\!+\!y^2}{N(y,y_{\scriptscriptstyle R})},\,\quad
P_2(y,y_{\scriptscriptstyle R},{\bf k}^2)=\qquad\quad\nonumber\\
\frac{(1-y^2_{\scriptscriptstyle R})/2+2y^2/3}{N(y,y_{\scriptscriptstyle R})}
-y^2_{\scriptscriptstyle R}
\left(\frac{1\!+\!y^2}{N(y,y_{\scriptscriptstyle R})}\right)^{\!\!2}
\frac{{\bf k}^2b^2}{9},\qquad
\label{a28a}
\en
with $N(y,y_{\scriptscriptstyle R})$ defined in Eq.~(\ref{a17b}).
The $R\to N+\pi$ and $R\to N+\sigma$ decay widths are defined by the matrix 
elements
\eq
{\cal M}^q_{R\to N+\pi}=3\langle\Psi_\pi,{\bf k}|\langle\Psi_N,-{\bf k}|
V^{eff}_q|\Psi_R,{\bf 0}\rangle\nonumber\\
=\frac{\sqrt{3}}{2}g^q_{\pi{\scriptscriptstyle NN}}p_2(y_\pi,{\bf k}^2)
\qquad\qquad\quad\label{a30}\\
\times e^{-\zeta(y_\pi)k^2b^2/6}
\langle{\scriptstyle\frac{1}{2}}{\scriptstyle S}^\prime_z|
{\bm\sigma}\cdot{\bf k}|
{\scriptstyle\frac{1}{2}}{\scriptstyle S}_z\rangle
\langle{\scriptstyle\frac{1}{2}}{\scriptstyle T}^\prime_z|\tau_{t_\pi}|
{\scriptstyle\frac{1}{2}}{\scriptstyle T}_z\rangle,\nonumber
\en
\eq
{\cal M}^q_{R\to N+\sigma}=3\langle\Psi_\sigma,{\bf k}|\langle\Psi_N,-{\bf k}|
V^{eff}_q|\Psi_R,{\bf 0}\rangle\nonumber\\
=\frac{\sqrt{3}}{2}g^q_{\sigma{\scriptscriptstyle NN}}p_4(y_\sigma,{\bf k}^2)
e^{-\zeta(y_\sigma){\bf k}^2b^2/6}
\delta_{{\scriptscriptstyle S}^\prime_z,{\scriptscriptstyle S}_z}
\delta_{{\scriptscriptstyle T}^\prime_z,{\scriptscriptstyle T}_z}
\label{a31}
\en
\medskip
with
\eq
p_4(y_\sigma,{\bf k}^2)=
-\frac{2}{3}\,\frac{1-y^2_\sigma}{n(y_\sigma)}
-\frac{{\bf k}^2b^2}{27}\qquad\qquad\qquad\\
\times\!\biggl[\!\frac{1\!+\!y^2_\sigma/2\!+\!y^4_\sigma/3}{n^2(y_\sigma)}
+\frac{{\bf k}^2b^2}{3}\frac{(1\!+\!y^2_\sigma)(1\!+\!y^2_\sigma/2)}{n^3(y_\sigma)}
\biggr].\nonumber
\label{a31a}
\en
As in the case of vector mesons the polynomials \\ 
$p_2(y_\pi,{\bf k}^2)$ and
$p_4(y_\sigma,{\bf k}^2)$ do not vanish in the limit \\ 
$|{\bf k}|\to\,$0 and the non-zero values
\eq
Z_\pi=\frac{2y_\pi^2/3}{n(y_\pi)},\quad
Z_\sigma=\frac{2}{3}\frac{1-y_\sigma^2}{n(y_\sigma)}
\label{a30z}
\en
determine the amplitudes (\ref{a30}) -- (\ref{a31}) for small values of 
$|{\bf k}|$.

\end{document}